\documentclass[%
aps,%
prl,%
reprint,%
amsmath,amssymb,%
floatfix,%
nofootinbib,%
superscriptaddress%
]{revtex4-2}

\usepackage[T1]{fontenc}
\usepackage[latin9]{inputenc}
\usepackage[english]{babel}
\usepackage{mathrsfs}
\usepackage{bbm}
\usepackage{graphicx}
\usepackage{xcolor}
\usepackage{physics}
\usepackage{bm}
\usepackage[export]{adjustbox}

\usepackage[colorlinks,
            linkcolor=blue,    
            citecolor=blue,  
            urlcolor=blue,
 	        bookmarks=true,        
	        bookmarksopen=true,    
	        bookmarksnumbered=true,
]{hyperref}

\newcommand{\id}{\mathbbm{1}}

\newcommand{\be}{\begin{equation}}
\newcommand{\ee}{\end{equation}}
\newcommand{\bea}{\begin{eqnarray}}
\newcommand{\eea}{\end{eqnarray}}

\begin{document}

\title{Quasi-local Edge Mode in XXX Spin Chain/Circuit with Interaction Boundary Defect}

\author{Toma\v z Prosen}

\affiliation{Department of Physics, Faculty of Mathematics and Physics, University of Ljubljana, Jadranska 19, SI-1000 Ljubljana, Slovenia}
\affiliation{Institute of Mathematics, Physics and Mechanics, Jadranska 19, SI-1000 Ljubljana, Slovenia}

\begin{abstract}
We study the Heisenberg spin-$1/2$ model on a semi-infinite chain --- or, equivalently, a trotterized unitary SU(2) symmetric six-vertex quantum circuit --- with a boundary defect where the interaction between the two spins nearest the edge differs from that in the bulk. For sufficiently strong boundary interaction we explicitly construct a conserved operator quasi-localized near the boundary using a matrix-product ansatz. This quasi-local edge mode leads to non-decaying boundary correlation functions, corresponding to a nonzero boundary Drude weight. The correlation length of the edge mode diverges at a finite critical value of the boundary interaction, signaling a transition to ergodic boundary dynamics for subcritical interactions.
\end{abstract}

\maketitle

\emph{Introduction.--}
 Breaking ergodicity in the dynamics of interacting quantum many-body systems is one of the most fascinating topics in contemporary research. While several general---albeit only partially understood---mechanisms are known, such as Yang-Baxter integrability~\cite{review16,review21}, strong quenched disorder~\cite{MBLbloch,MBLvidmar}, and kinematic constraints~\cite{Garrahan20,MarinoPRXQ22,BertiniPRL24}, they share a common thread: ergodicity breaking is generally tied to the existence of nontrivial local or quasi-local conserved operators.
Whereas most studies have focused on conserved operators supported in the bulk, such operators can instead originate from system boundaries. Integrable spin chains can---and generically do---host conserved operators (quasi-)localized near the edges~\cite{FendleyPRB14,FendleyJPA16,Fendley25,Gehrmann26,FendleyPRX17,Kemp25}, known as strong zero modes (SZMs), which can be understood as interacting analogues of Majorana edge modes of quasi-free fermionic chains. All SZM constructions known to date require either anisotropic interactions or boundary fields (and are generic for large anisotropies~\cite{Gehrmann26b}).

The concepts of Yang-Baxter integrability~\cite{VanicatPRL18}, quasi-local conserved charges~\cite{LjubotinaPRL19}, and SZMs~\cite{VernierPRL24} have recently been extended to (Floquet) quantum circuits, which provide a natural framework for studying discrete-time, locally interacting quantum many-body dynamics, both theoretically and experimentally in the context of digital quantum simulation~(see e.g.~\cite{GQAI_Nature22,GQAI_Science24a,GQAI_Science24b}).

A complementary, and until now disconnected, line of research concerns spin impurities attached to the boundary of integrable chains. The isotropic Heisenberg chain whose edge bond differs from the bulk exchange---equivalently, a spin-$1/2$ impurity exchange-coupled to the end of a semi-infinite chain---admits an exact Bethe-ansatz solution~\cite{Frahm97}. Revisiting this problem, Refs.~\cite{Kattel24,Kattel25b} recently uncovered a \emph{boundary eigenstate phase transition} at the critical ratio $4/3$ of boundary to bulk exchange, separating a Kondo phase, where the edge spin is screened by a multiparticle Kondo cloud, from a bound-mode phase, where it is screened by a single-particle mode exponentially localized at the edge and the spectrum reorganizes into disconnected towers; analogous transitions arise for anisotropic interactions~\cite{KattelXXZ} and in noisy integrable impurity circuits~\cite{Kattel25a}. These remarkable results are, however, statements about individual eigenstates and equilibrium thermodynamics; their consequences for dynamics, in particular for relaxation at high temperature, have remained open.

In this Letter, we connect the two strands. We propose a simple modification of a well-known integrable quantum circuit --- a staircase version of the trotterized SU(2)-symmetric spin-$1/2$ Heisenberg chain --- in which the interaction strength between the first two spins at one boundary is varied (equivalently, we consider a half-infinite chain with an edge bond defect). We show that, above a critical boundary interaction strength, the system hosts an exactly conserved quasi-local edge mode, constructed in closed form in terms of a $16\times16$ matrix-product ansatz with no free (spectral) parameter, at any point of the two-parameter circuit --- away from the Trotter limit, where no Bethe-ansatz solution is available. The edge mode yields, via a saturated Mazur bound, an explicit infinite-temperature boundary autocorrelator (boundary Drude weight), vanishing at a critical line where the localization length of the mode diverges. In the continuous-time limit the critical coupling is $g_{\rm c}=4/3$, precisely the boundary eigenstate transition of Refs.~\cite{Kattel24,Kattel25b}: our quasi-local edge mode can thus be understood as the operator-space avatar of the boundary bound mode, promoting an eigenstate-level transition to a transition between non-ergodic and ergodic boundary dynamics at infinite temperature. The construction itself is unexpected on two counts. First, in contrast to all known (quasi-local) SZMs~\cite{FendleyPRB14,FendleyJPA16,Fendley25,Gehrmann26,FendleyPRX17,VernierPRL24,Kemp25}, it requires neither anisotropy nor boundary fields, operating in a fully SU(2)-symmetric setting. Second, although the continuous-time limit is Bethe-ansatz integrable~\cite{Frahm97,Kattel24}, the relevance of the standard quantum inverse scattering method for our construction appears highly nontrivial, as it would require a 16-dimensional auxiliary space.

\begin{figure}[tbp]
	\centering
	\includegraphics[width=0.75\columnwidth]{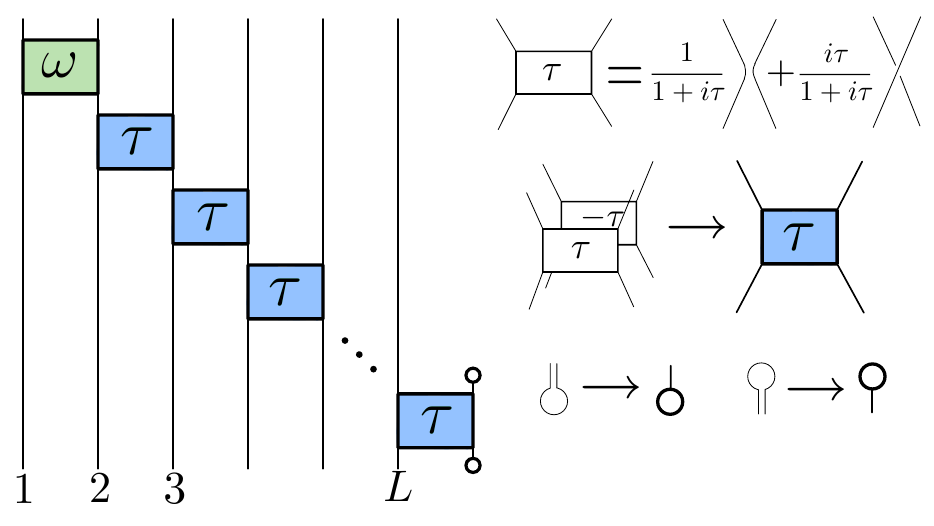}
	\caption{Schema of the six vertex staircase circuit with boundary interaction defect. On the last qubit ($L$) we apply depolarizing channel obtained by placing a virtual identity on site $L+1$ and then tracing out that site.}
	\label{fig:scheme}
\end{figure}

\emph{Heisenberg circuit on a half-chain with boundary interaction defect.--} Consider a chain of $L$ qubits (spins $1/2$) over Hilbert space ${\mathcal H}_L=(\mathbb C^2)^{\otimes L}$ and local Pauli operators $\sigma^{\nu}_n$, $\nu\in\{0,1,2,3\}$, $n\in\{1,2\ldots L\}$ generating ${\rm End}(\mathcal H_L)$.
We write a unitary SU(2)-symmetric 6-vertex (or XXX) gate as
\be
U_{n,n+1}(\tau)=(\id + {\rm i}\tau P_{n,n+1})/(1+{\rm i}\tau),\quad \tau\in\mathbb R,
\ee
where $P_{n,n+1}=\frac{1}{2}\sum_\nu \sigma^\nu_n \sigma^\nu_{n+1}$ exchanges qubits $n$, $n+1$. 
We will be interested in the limit of infinite half-chain $L\to\infty$, but we may
keep $L$ finite for comparison with numerical simulations.

We consider a staircase Floquet XXX circuit generated by
\be
\mathcal U_{[L]} =U_{L-1,L}(\tau)\cdots U_{34}(\tau)U_{23}(\tau)U_{12}(\omega)
\ee
where the boundary interaction $\omega\in\mathbb R$ may differ from the bulk coupling $\tau$. Writing $\omega = g\tau$,
the Trotter limit $\tau\to 0$ yields $\mathcal U \simeq \exp(-{\rm i}\tau H)$ with Hamiltonian $H =-g P_{12} - \sum_{n=2}^{L-1} P_{n,n+1}$.
The dynamics of Hermitian observables $Q$, $Q^{(t)} = \mathcal U^t_{[L]} Q \mathcal U^{-t}_{[L]}$, can be represented in terms of dynamics of $4^L$ dimensional real Pauli vectors 
$q_{\underline{\nu}} = (\sigma^{\underline{\nu}}|Q)$, $\sigma^{\underline{\nu}}=\sigma^{\nu_1}_1\sigma^{\nu_2}_2\cdots \sigma^{\nu_L}_L$, $(X|Y) = 2^{-L} \tr X^\dagger Y$ :
\be
q^{(t)} = \mathcal W_{[L]}^t q
\ee
where the Floquet-Pauli-circuit propagator
\be
\mathcal W_{[L]} = W_{12}(\omega)W_{23}(\tau)W_{34}(\tau) \cdots W_{L-1,L}(\tau)
\label{W}
\ee
is generated by $16\times 16$ real orthogonal matrices $W_{\nu_1\nu_2,\nu'_1\nu'_2}(\tau)=\frac{1}{4}\tr(\sigma^{\nu_1}\otimes \sigma^{\nu_2})U(\tau) 
(\sigma^{\nu'_1}\otimes \sigma^{\nu'_2}) U(-\tau)$, App.~Eq.~(\ref{eq:W}), $W_{n,n+1} = \id_{4^{n-1}}\!\otimes\!W\!\otimes\!\id_{4^{L-n-1}}$.
Let $|\underline{\nu}| = {\rm max}\{n|\nu_n>0\}$ represent the length of a Pauli string $\underline{\nu}$.
We define the partial Hilbert-Schmidt (HS) norms, or the length distribution of an operator:
\be
\mathcal P_\ell(Q) = \sum_{\underline{\nu}} \delta_{\ell,|\underline\nu|} |q_{\underline{\nu}}|^2,
\ee
satisfying the sum rule $\sum_\ell \mathcal P_\ell(Q) = (Q|Q) = \|Q \|^2_{\rm HS}$. 

\noindent {\em Definition:} Operator $Q \in {\rm End}(\mathcal H_L) $ is {\em boundary quasi-local} if $\exists c,\gamma>0$, s.t. $\mathcal P_{\ell}(Q) < c e^{-\gamma\ell}$ uniformly in $L$.

In this Letter we will be exploring existence of a boundary quasi-local conserved charge that commutes with $\mathcal U$ --- quasi-local edge mode (QLEM).

\begin{figure}[tbp]
	\centering
	\includegraphics[width=0.54\columnwidth]{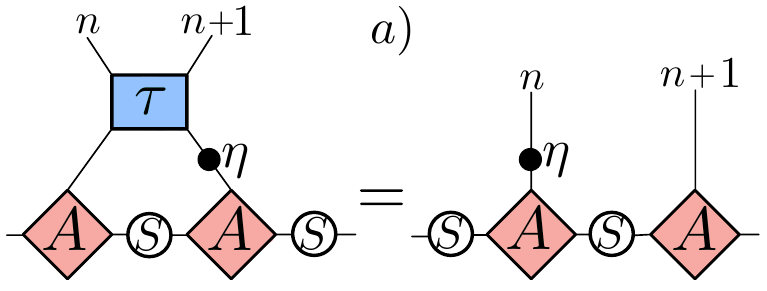}\quad\includegraphics[width=0.42\columnwidth]{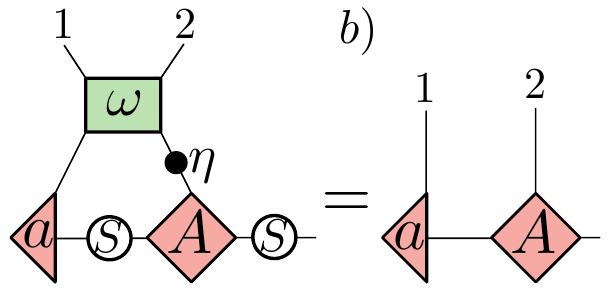}
	\vspace{1mm}

	\includegraphics[width=0.73\columnwidth]{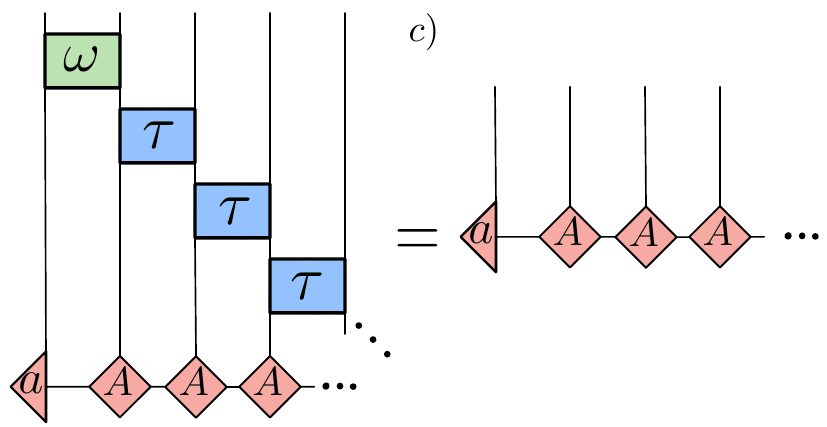}
	
	\vspace{-2mm}

	\caption{Cancelation algebra diagrams (a): Eq. (\ref{bulk}), (b): Eq. (\ref{boundary}),  and their implementation to demonstrate fixed point (conservation law) condition (c) for the boundary matrix product ansatz (\ref{MPA}).}
	\label{fig:cancellation}
\end{figure}

\begin{figure}[tbp]
	\centering
	\includegraphics[width=0.96\columnwidth]{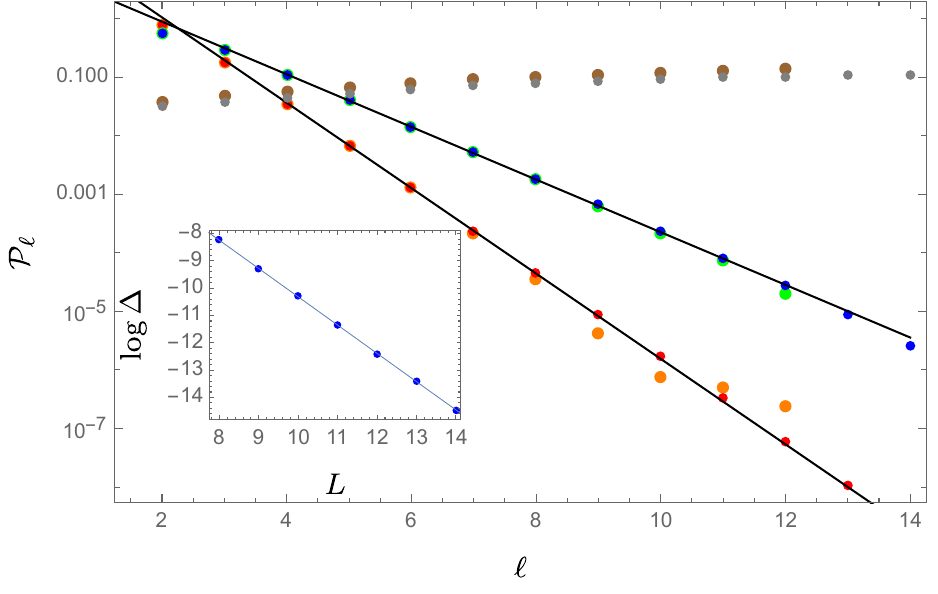}
	\vspace{-2mm}
	\caption{Partial norm profiles $\mathcal P_{\ell}$ of the subleading eigenoperator $Q^{(1)}$ of the staircase channel (Fig.~\ref{fig:scheme}),
	obtained with quasi-exact numerics for $\omega=\tan 0.9$, and for $\tau = \tan 0.4$: orange (red), $\tau = \tan 0.5$: green (blue), $\tau = \tan 1.3$ brown (gray) dots,
	for $L=12$ ($L=14$). Straight lines indicate analytical results: exponentials $\propto\xi^\ell$ where
	$\xi=0.18667$ and $\xi= 0.35529$, respectively, are the spectral gaps $\xi=\Lambda_1/\Lambda_0$ of the transfer matrix (\ref{HST}).
	The inset shows the spectral gap $\Delta$ of the channel $\mathcal M_{[L]}$ vs. $L$ (dots) compared to $\xi^L$ (line).}
	\label{fig:numerics}
\end{figure}

{\em QLEM from boundary dissipated channel.--} In order to target QLEM in finite systems we will consider a Heisenberg-picture 
depolarization channel
\be
\mathcal M(Q) = \frac{1}{2}\tr_{L+1} \mathcal U_{[L+1]} (Q\otimes \id_{2}) \mathcal U^{-1}_{[L+1]}
\ee
 with Pauli basis matrix representation
$\mathcal M_{[L]} = \mathcal{W}_{[L]} w_L$, $\mathcal W_{[L]}$ from Eq.~\eqref{W}, where $w_L$ is matrix representation of a single qubit depolarization channel with matrix elements
$w_{\nu,\nu'} = W_{\nu0,\nu'0}(\tau)$ acting on the right-most qubit. The channel $\mathcal M$ is designed to dissipate (continously decimate) long operators, i.e. those that have non-identity component on the right-most site $n=L$.
Let us consider eigenvectors, $\mathcal M_{[L]} q^{(m)} = \mu_m q^{(m)}$, with eigenvalues $\mu_m$ of largest moduli $|\mu_m|\le 1$, projecting out trivial unit eigenoperator with $\mu_0=1$.
We find excellent numerical evidence that in certain parameter regime $\omega > \omega_{\rm c}(\tau)$ the subleading eigenoperator, which we obtain by a simple iteration $q^{(1)}= \lim_{t\to\infty} 
\mathcal M_{[L]}^t b/\| \mathcal M_{[L]}^t b\|$, where $b$ encodes a relevant initial (traceless) local boundary operator, say $\vec{\sigma}_1\cdot\vec{\sigma}_2$,
behaves as QLEM; see Fig.~\ref{fig:numerics}. If the operator $Q$ were exactly conserved in the thermodynamic limit $L\to\infty$, then under the assumption of exponential localization of partial norms,
the longest $\ell=L$ Pauli terms in $Q$ have total HS weight $\mathcal P_L(Q)\propto e^{-\gamma L}$. These are the only terms affected by non-unitarity of the channel $\mathcal M_{[L]}$, hence first order
perturbation theory implies the exact same scaling of the Liouvillian spectral gap
\be
\Delta=1-\mu_1 \propto e^{-\gamma L}.
\ee
Performing quasi-exact numerics for $L\le 14$, fixing $\omega$ and varying $\tau$ we found very strong hints on the existence of QLEM for $\tau < \tau_{\rm c}$ (see Fig.~\ref{fig:numerics}).
Furthermore, exploring the Schmidt singular value spectra of vectors $q^{(1)}_{\underline{\nu}}$ with multi index $\underline{\nu}$ bi-partitioned to $\nu_1,\ldots \nu_N$ and $\nu_{N+1},\ldots \nu_L$ we found that, while fixing $N$ and increasing $L$, the spectra had $16$ dominant (and frozen) singular values while the remaining singular values were decreasing exponentially with $L$. Furthermore, by increasing $N$ for sufficiently large $L$, the `frozen' $16$ singular values were further decreasing, exponentially in $N$ (except the leading one approaching $1$).
This was further corroborated by time-evolving block decimation (TEBD) simulation of $\mathcal M_{[L]}^t b$ using a matrix product ansatz (MPA) of fixed bond dimension $16$
having truncation error which became negligible for sufficiently large $L$.
The MPA fixed point $q^{(1)}$ was further polished removing all freedom allowed by local gauge invariance of MPA, while finally 
a unique analytical translationally invariant MPA form of QLEM has been guessed which is reported and proven in the following paragraph.

\begin{figure}[tbp]
	\centering
	\includegraphics[width=0.96\columnwidth]{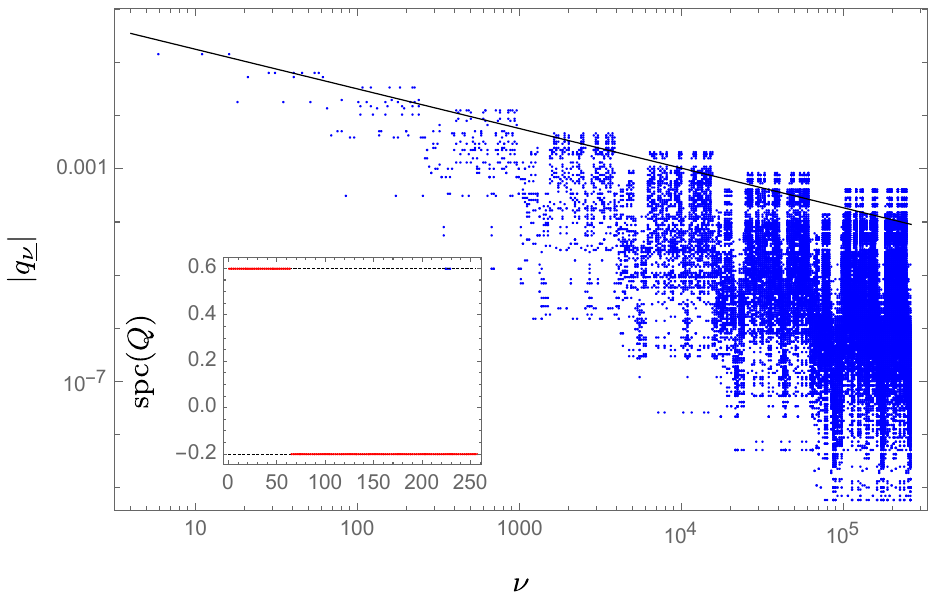}
	\vspace{-2mm}
	\caption{QLEM vector components $q_{\underline{\nu}}$ vs. integer ordered basis 
	$\nu=\sum_{n=1}^L \nu_n 4^{n-1}$ for $\omega=\tan 0.9$, $\tau=\tan 0.5$
	computed from analytical solution (\ref{MPA}), evaluated up to $L=9$.
	Black line indicates partial norm scaling $\nu^{\log_4\!\xi}$ from the spectral gap $\xi= 0.35529$. Inset: 256 eigenvalues of $Q$ for $L=8$, compared to analytical predictions $\|Q\|_{\rm HS}\sqrt{3}$,$-\|Q\|_{\rm HS}/\sqrt{3}$ (dashed). }
	\label{fig:q}
\end{figure}

{\em Analytic matrix product structure of QLEM and phase transition.--} We postulate existence of {\em real}, $\omega,\tau$ dependent matrices $A^\nu$, $\nu=0,1,2,3$, 
and boundary vectors $\bra{a^\nu}$, $\ket{\rm r}$, such that QLEM $Q=\sum_{\underline{\nu}}  q_{\underline{\nu}} \sigma^{\underline{\nu}}$ 
would be given by the following MPA, for arbitrary $L$:
\be
q_{\underline{\nu}} = \bra{a^{\nu_1}} \lambda_0^{-1}A^{\nu_2} \lambda_0^{-1}A^{\nu_3}  \lambda_0^{-1}A^{\nu_4}\cdots\ket{\rm r}\,,
\label{MPA}
\ee
where $\ket{\rm r}$ is an eigenvector of $A^0$ of maximal modulus eigenvalue $\lambda_0$, $A^0\ket{\rm r} = \lambda_0\ket{\rm r}$.
We claim that $q$ satisfies fixed point (conservation) condition $(W_{12} W_{23} \cdots)q = q$ in the limit $L\to\infty$ if the MPA matrices satisfy the bulk equation
\be
\sum_{\alpha\alpha'}W_{\nu\nu',\alpha\alpha'}(\tau) \eta_{\alpha'} A^{\alpha} S A^{\alpha'} S = \eta_{\nu} S A^\nu S A^{\nu'}\,,
\label{bulk}
\ee
and the left boundary equation (diagramatics in Fig.~\ref{fig:cancellation})
\be
\sum_{\alpha\alpha'}W_{\nu\nu',\alpha\alpha'}(\omega) \bra{a^\alpha} S A^{\alpha'} S = \bra{a^{\nu}} A^{\nu'},
\label{boundary}
\ee
where $S$ is an involution in auxiliary space $S^2 = \id$, and similarly $\eta^2_{\nu}=1$. 
Importantly, we do not require an additional right boundary cancellation condition,
as it should come for free in the asymptotic regime $L\to\infty$ under the assumption of QLEM. 
In other words, assuming exponentially small weights of long Pauli components 
$q_{\underline{\nu}}$, one can replace $\cdots A^{\nu_n}A^{\nu_{n+1}}\cdots$
by $\cdots A^{\nu_n} S \eta_{\nu_{n+1}} A^{\nu_{n+1}} S \cdots$ 
at exponentially small in $n$ cost in HS norm (i.e., far enough to the right).

A solution of (\ref{bulk},\ref{boundary}) exists over 16 dimensional auxiliary space $\mathcal V=\mathbb R^{16}$, 
where nonzero entries can be encoded in a set of $21$ variables $\{x_i\}_{i=1}^{21}$, Eqs.~(\ref{eq:a},\ref{eq:AA}) in App., which is minimal in a sense that there is no
linear relation among $x_i$ with constant coefficients. Remarkably, {\em all} entries of the solution can be expressed in terms of a single square-root radical,
namely $x_i = u_i + r v_i$ where $u_i(\omega,\tau),v_i(\omega,\tau)$ are rational functions and
\be
r(\omega,\tau)= \sqrt{\omega ^3 \left(\left(4-\tau ^2\right) \omega -4 \tau \right)}\,.
\ee
Cartesian components $\nu=1,2,3$ are intertwined 
\be
\bra{a_{\nu'}} = \bra{a_{\nu}} P,\quad
A^{\nu'} = P^{-1} A^{\nu} P, \quad \nu' = {\rm mod}(\nu,3)+1
\ee
by a cyclic permutation operator $P$ over $\mathcal V$,  $P^3=\id$:
\bea
P&=&\ket{5}\!\bra{1}+
\ket{6}\!\bra{2}
+\ket{1}\!\bra{3}
+\ket{2}\!\bra{4}
+\ket{3}\!\bra{5}
+\ket{4}\!\bra{6} \nonumber \\
&+& \ket{11}\!\bra{7}\!+\!\ket{12}\!\bra{8}\!+\!\ket{7}\!\bra{9}\!+\!\ket{8}\!\bra{10}\!+\!\ket{9}\!\bra{11}\!+\!\ket{10}\!\bra{12} \nonumber \\
&+&\ket{15}\!\bra{13} +\ket{13}\!\bra{14} +\ket{14}\!\bra{15} +\ket{16}\!\bra{16},
\eea
where $\{\ket{i}\}_{i=1}^{16}$ denotes canonical basis of $\mathcal V$.
Both, auxiliary and physical space involutions are diagonal, reading
\be
S=(-\id_6)\oplus (\id_{10}),\quad \eta = (\id_1)\oplus(-\id_3)\,.
\ee
Furthermore, this analytic solution appears to be unique and has no free 
parameter (analog to the spectral parameter in integrability structures).

\begin{figure}[tbp]
	\centering
	
	\vspace{1mm}
	\includegraphics[width=0.98\columnwidth]{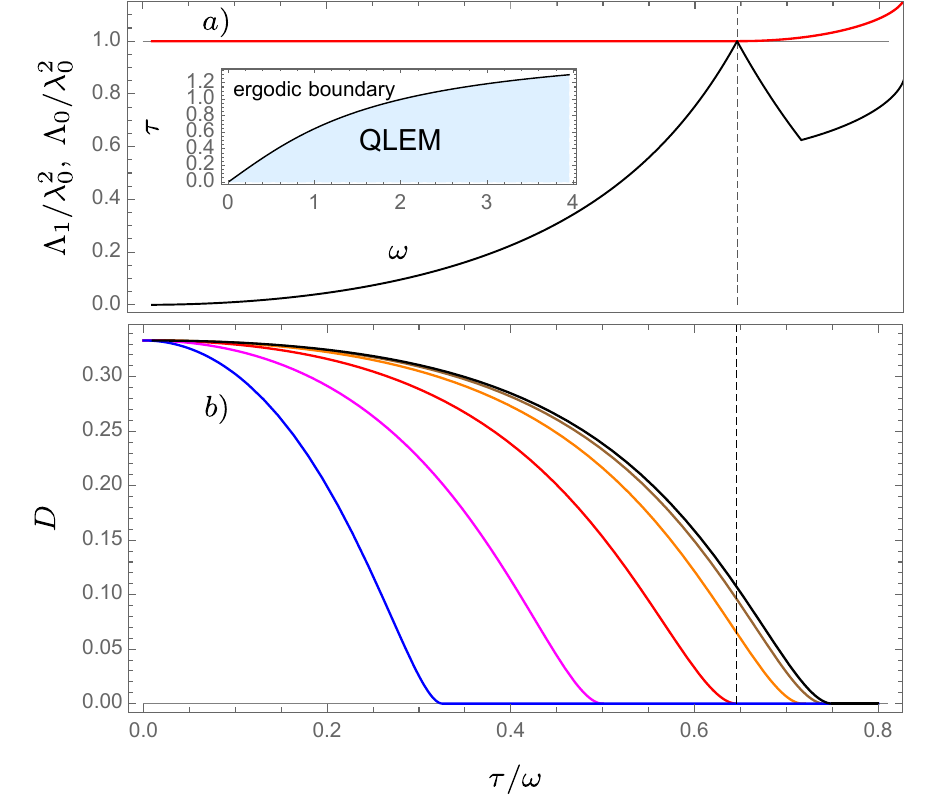}
	\caption{(a) The relative leading and subleading eigenvalues of the HS transfer matrix $\mathbb T$ (\ref{HST}), $\Lambda_0/\lambda_0^2$ (red), $\Lambda_1/\lambda_0^2$ (black),
	vs. $\tau$ for $\omega=1$. The inset gives a $(\omega,\tau)$phase-diagram with the region of existence of QLEM (where 
	$\Lambda_0=\lambda_0^2$) indicated in light-blue. (b) Boundary Drude weights vs $\tau/\omega$ for $\omega=4$ (blue), $2$ (magenta), $1$ (red), $1/2$ (orange), $1/4$ (brown), and
	countinuous time case $\omega \to 0$ (black). Dashed vertical line denotes the critical point when the gap $1-\xi$ closes for $\omega=1$. }
	\label{fig:Drude}
\end{figure}

In analogy to quasi-locality in translationally invariant systems~\cite{prosen13,ilievski15,review16}, one can characterize
quasi-locality of the edge mode by the $L\to\infty$ convergence of HS norm of MPA (\ref{MPA}) while keeping the short Pauli components fixed. 
This statement can be formalized by defining HS transfer matrix
\be
\mathbb T = \sum_{\nu=0}^3 A^\nu \otimes A^\nu\,,
\label{HST}
\ee
and denoting its leading and subleading (in modulus) eigenvalues, respectively, as $\Lambda_0$, $\Lambda_1$: Operator (\ref{MPA}) is quasi-local if $\Lambda_0=\lambda_0^2$, while $|\Lambda_1|<\Lambda_0$. Then the partial-norm exponent 
 is given by the spectral gap \be
\mathcal P_\ell \propto e^{-\gamma \ell},\quad
 \gamma = -\log|\xi|, \quad \xi=\Lambda_1/\Lambda_0.
 \ee
We found a relevant eigenvalue $\lambda_*$ and right eigenvector $\ket{\rm r_*}$ of $A^0$, $A_0\ket{\rm r_*}=\lambda_*\ket{\rm r_*}$:
\bea
&&\lambda_* \!=\! \frac{\tau ^4\!+\!\tau ^3 \omega\!+\!5 \left(\tau ^2\!\!+\!2\right) \omega ^4\!-\!\left(8 \tau ^2\!\!+\!9\right) \tau  \omega ^3\!+\!3 \left(\tau ^2\!\!-\!1\right)
   \tau ^2 \omega ^2}{\tau ^2 \left(\omega ^2+1\right) \left(\left(\tau ^2+5\right) \omega ^2+\tau ^2-2 \tau  \omega \right)} \nonumber \\
&&\quad+ r\left(\frac{\tau ^2+1}{\tau ^2 \left(\omega ^2+1\right)}-\frac{\tau ^2+4}{\left(\tau ^2+5\right) \omega ^2+\tau ^2-2 \tau  \omega }\right)\,, \!\!\!\!\!\!\!\!\\
&&\ket{\mathrm r_*} = \frac{1}{\sqrt{3}}\left(\ket{13}+\ket{14}+\ket{15}\right),
\eea
which, as verified by direct computation, correspond as well to a left eigenvector of (\ref{HST}) of eigenvalue $\lambda_*^2$: 
\be
\bra{\rm r_*}\!\otimes\!\bra{\rm r_*}\mathbb T = \lambda_*^2 \bra{\rm r_*}\!\otimes\!\bra{\rm r_*}.
\ee 
It can be demonstrated (see Fig.~\ref{fig:LF} in App.) that this is exactly the leading eigenpair $\lambda_0=\lambda_*$, $\ket{\rm r}=\ket{\rm r_*}$ in the QLEM region $\Lambda_0=\lambda_0^2$.
The corresponding right eigenvector $\mathbb T\ket{\rm R}=\lambda_0^2 \ket{\rm R}$ is reported in App., Eq.~(\ref{RR}), while for
computation of the spectral gap $\xi$ we have to resort to numerics.
In Fig.~\ref{fig:q} we plot operator components $q_{\underline{\nu}}$ 
of a typical QLEM in log-log scale as computed from analytic MPA (\ref{MPA}) in integer indexed basis $\nu=\sum_n 4^{n-1}\nu_n$ in comparison with the predicted overall scaling $\sim\nu^{\log_4\!\xi}$.
We illustrate the behavior of $\Lambda_{0,1}/\lambda_0^2$ on parameters $\omega,\tau$ in Fig.~\ref{fig:Drude}a and sketch a phase-diagram of region where $\Lambda_0=\lambda_0^2$.
We note that if one approaches a critical line, say at fixed $\omega$, then $\log\xi \propto \tau-\tau_{\rm c}(\omega)$ so the {\em correlation length} scales with critical exponent $-1$,
$1/\gamma \propto |\tau-\tau_{\rm c}|^{-1}$.

{\em Boundary Drude weight.--} Consider a traceless operator $B=\sum_{\underline{\nu}} b_{\underline{\nu}} \sigma^{\underline{\nu}}$
localized near the edge. Boundary ergodicity is characterized by averaged time autocorrelation function, dubbed Drude weight
\be
D=\lim_{T\to\infty}\lim_{L\to\infty}\frac{1}{T}\sum_{t=1}^T \bigl(B\vert\mathcal U^t_{[L]}B\mathcal U^{-t}_{[L]}\bigr).
\ee
Existence of an QLEM $Q$ immediately provides a Mazur lower bound on $D$,
$D \ge |(B|Q)|^2/(Q|Q)$, a signature of non-ergodic (or non-relaxing) boundary dynamics. 
Moreover, since our analytic MPA (\ref{MPA}) is unique and reproduces exact and TEBD numerics for various specific choices of initial boundary operator (see e.g. Fig.~\ref{fig:numerics}), 
we conjecture that QLEM -- when exists -- is unique and does not depend on any additional relevant parameter. Thus the Mazur bound has to saturate and we can write explicit expression 
for the Drude weight
\be
D= \frac{(\bra{\rm r_*}\!\otimes\!\bra{\rm r_*})\ket{\rm R}\left|\sum_{\underline{\nu}\in\mathbb Z_4^{\beta}} b_{\underline{\nu}} \bra{a^{\nu_1}}A^{\nu_2}\cdots A^{\nu_{\beta}}\ket{\rm r_*}\right|^2}{\lambda_*^{2(\beta-1)}\Bigl(\sum_{\nu\in\mathbb Z_4} (\bra{a^\nu}\!\otimes\!\bra{a^\nu}\Bigr)\ket{\rm R}}\,,
\label{Drude}
\ee
where $\beta$ is the Pauli length of operator $B$.
Choosing normalized boundary interaction $B=\frac{1}{\sqrt{3}}\vec{\sigma}_1\cdot\vec{\sigma}_2$ we find an explicit expression, Eq. (\ref{eq:Drude1}) in App., which is valid
as long as $\Lambda_0=\lambda_0^2$. It turns out that analytic expression of Drude weight (\ref{eq:Drude1}) can be used to determine the critical line $\tau=\tau_{\rm c}(\omega)$ by the condition of minimum $D(\omega,\tau)=0$.
In the limit of continous time boundary-Heisenberg Hamiltonian, $\omega=g\tau$, $\tau\to 0$, we find QLEM to exist above the critical coupling $g > g_{\rm c}=4/3$ (cf. Refs.\cite{Kattel24, Kattel25b}). See Fig.~\ref{fig:Drude}b.
We note that the transfer matrix eigenvalues and Drude weight are even functions, e.g $D(\omega,\tau) = D(-\omega,-\tau)$, hence for generally signed $\omega,\tau$, the
region of existence of QLEM and positivity of $D$ can be extended to $-\tau_c(-\omega) \le \tau \le \tau_c(\omega)$.

{\em Discussion.--}
We have uncovered a simple algebraic matrix-product structure in the XXX spin-$1/2$ chain, or equivalently in the trotterized XXX qubit circuit, with an interaction defect near the boundary. This structure yields a conserved quasi-local edge operator above a critical boundary interaction strength. 
The deeper relation of our construction to strong zero modes $\Psi$ introduced and studied by Fendley and collaborators~\cite{FendleyJPA16,FendleyPRB14,VernierPRL24,Fendley25}, and to their
recent reformulation in terms of commutant algebras~\cite{Motrunich26}, remains unclear. 
In contrast to all known SZMs, which satisfy $\Psi^2=\id$ and $\tr\Psi=0$, our QLEM $Q$ appears to obey the more general quadratic relation
\be
Q^2 - 2 c Q = 3 c^2\id,\quad c = \sqrt{(Q|Q)/3},\quad \tr Q = 0,
\label{eq:Q2}
\ee
so that one quarter of its eigenvalues equal $3 c$ and three quarters equal $-c$; see Fig.~\ref{fig:q}(inset), where $(Q|Q)$ can be computed via (\ref{HST}), see App. Eq.~(\ref{eq:QQ}).
The parametric existence of operator $Q$ implies a transition between ergodic and non-ergodic dynamics for observables localized near the boundary.

 In the continuous-time limit, the transition point coincides with the
 boundary eigenstate (Kondo-to-bound-mode) transition identified via
 Bethe ansatz~\cite{Kattel24,Kattel25b}; it would be very interesting to
 understand the QLEM directly in terms of the boundary-string solutions
 of the Bethe equations, and conversely, whether the spectral splitting
 of $Q$ into $3c$ and $-c$ sectors, Eq.~(\ref{eq:Q2}), reflects the
 tower structure of the Hilbert space found in~\cite{Kattel24,Kattel25b}.

The central algebraic relation~(\ref{bulk}) can be interpreted as a `decorated' version of RLL or Yang-Baxter relation, 
which lies at the heart of quantum integrability. However, the precise role of the delimiter operators $S$ and $\eta$ in this auxiliary quantum inverse scattering problem remains to be clarified~\cite{Note1}.
Furthermore, it is important to clarify the role of SU(2) symmetry---which appears to be unique to our setup compared to related literature---and whether our construction can be deformed to $U(1)$-symmetric XXZ interactions.
Conceptually, our matrix-product structure is reminiscent of a recently discovered matrix-product operator in the XYZ chain~\cite{FendleyXYZ25}. A crucial difference, however, is that our construction contains no spectral parameter, rendering the quasi-local conserved charge unique and thus single-handedly controlling Drude weights.
We expect these results to open a new path for exploring boundary-induced dynamical phenomena in interacting quantum many-body systems.

  The author warmly thanks F. Essler and P. Fendley for fruitful discussions.
  Support by Advanced Grant No.~101096208 -- QUEST of European Research Council (ERC), 
  and Research Programme P1-0402 and grant N1-0368 of Slovenian Research and Innovation Agency 
  (ARIS) is greatfully acknowledged.
  
\bibliography{bibfile.bib}

@misc{Note1,
 note = {Besides the staircase circuit geometry studied above, we also have an equivalent QLEM $Q=\sum_{\underline{\nu}} q_{\underline{\nu}} \sigma^{\underline{\nu}}$ 
for the  so-called brickwork circuit with boundary interaction deffect $\mathcal W_{\rm bw}=\mathcal W_{\rm o} \mathcal W_{\rm e}$, 
$\mathcal W_{\rm e} = W_{23}(\tau) W_{45}(\tau) \cdots$, $\mathcal W_{\rm o} = W_{12}(\omega) W_{34}(\tau) W_{56}(\tau)\cdots$, namely $q_{\underline{\nu}} = \bra{a^{\nu_1}} B^{\nu_2} C^{\nu_3} B^{\nu_4} C^{\nu 5}\cdots\ket{\rm r}$, where $B^{\nu} = \lambda^{-1}_0 A^{\nu} S$, $C^{\nu} = \eta_\nu B^{\nu}$ where Eqs.~(\ref{bulk},\ref{boundary}) now imply $\mathcal W_{\rm bw}q=q$.}
}

@article{BertiniPRL24,
  title = {Localized Dynamics in the Floquet Quantum East Model},
  author = {Bertini, Bruno and Kos, Pavel and Prosen, Tomaz},
  journal = {Phys. Rev. Lett.},
  volume = {132},
  issue = {8},
  pages = {080401},
  numpages = {7},
  year = {2024},
  month = {Feb},
  publisher = {American Physical Society},
  doi = {10.1103/PhysRevLett.132.080401},
  url = {https://link.aps.org/doi/10.1103/PhysRevLett.132.080401}
}

@article{Frahm97,
  title={The open spin chain with impurity: an exact solution},
  author={Frahm, Holger and Zvyagin, Andrei A},
  journal={Journal of Physics: Condensed Matter},
  volume={9},
  number={45},
  pages={9939--9946},
  year={1997},
  url={https://iopscience.iop.org/article/10.1088/0953-8984/9/45/021/meta}
}

@misc{Motrunich26,
      title={Strong Zero Modes via Commutant Algebras}, 
      author={Sanjay Moudgalya and Olexei I. Motrunich},
      year={2026},
      eprint={2603.02326},
      archivePrefix={arXiv},
      primaryClass={cond-mat.stat-mech},
      url={https://arxiv.org/abs/2603.02326}, 
}

@misc{Gehrmann26,
      title={Exact strong zero modes in quantum circuits and spin chains with non-diagonal boundary conditions}, 
      author={Sascha Gehrmann and Fabian H. L. Essler},
      year={2026},
      eprint={2511.05490},
      archivePrefix={arXiv},
      primaryClass={cond-mat.stat-mech},
      url={https://arxiv.org/abs/2511.05490}, 
}

@article{review21,
  title = {Finite-temperature transport in one-dimensional quantum lattice models},
  author = {Bertini, B. and Heidrich-Meisner, F. and Karrasch, C. and Prosen, T. and Steinigeweg, R. and \ifmmode \check{Z}\else \v{Z}\fi{}nidari\ifmmode \check{c}\else \v{c}\fi{}, M.},
  journal = {Rev. Mod. Phys.},
  volume = {93},
  issue = {2},
  pages = {025003},
  numpages = {71},
  year = {2021},
  month = {May},
  publisher = {American Physical Society},
  doi = {10.1103/RevModPhys.93.025003},
  url = {https://link.aps.org/doi/10.1103/RevModPhys.93.025003}
}

@article{MBLbloch,
  title = {Colloquium: Many-body localization, thermalization, and entanglement},
  author = {Abanin, Dmitry A. and Altman, Ehud and Bloch, Immanuel and Serbyn, Maksym},
  journal = {Rev. Mod. Phys.},
  volume = {91},
  issue = {2},
  pages = {021001},
  numpages = {26},
  year = {2019},
  month = {May},
  publisher = {American Physical Society},
  doi = {10.1103/RevModPhys.91.021001},
  url = {https://link.aps.org/doi/10.1103/RevModPhys.91.021001}
}

@article{MBLvidmar,
  title={Many-body localization in the age of classical computing},
  author={Sierant, Piotr and Lewenstein, Maciej and Scardicchio, Antonello and Vidmar, Lev and Zakrzewski, Jakub},
  journal={Reports on Progress in Physics},
  volume={88},
  number={2},
  pages={026502},
  year={2025},
  url={https://iopscience.iop.org/article/10.1088/1361-6633/ad9756/meta},
  publisher={IOP Publishing}
}

@article{Garrahan20,
  title = {Quantum East Model: Localization, Nonthermal Eigenstates, and Slow Dynamics},
  author = {Pancotti, Nicola and Giudice, Giacomo and Cirac, J. Ignacio and Garrahan, Juan P. and Ba\~nuls, Mari Carmen},
  journal = {Phys. Rev. X},
  volume = {10},
  issue = {2},
  pages = {021051},
  numpages = {21},
  year = {2020},
  month = {Jun},
  publisher = {American Physical Society},
  doi = {10.1103/PhysRevX.10.021051},
  url = {https://link.aps.org/doi/10.1103/PhysRevX.10.021051}
}

@article{MarinoPRXQ22,
  title = {Kinetically Constrained Quantum Dynamics in Superconducting Circuits},
  author = {Valencia-Tortora, Riccardo J. and Pancotti, Nicola and Marino, Jamir},
  journal = {PRX Quantum},
  volume = {3},
  issue = {2},
  pages = {020346},
  numpages = {27},
  year = {2022},
  month = {Jun},
  publisher = {American Physical Society},
  doi = {10.1103/PRXQuantum.3.020346},
  url = {https://link.aps.org/doi/10.1103/PRXQuantum.3.020346}
}

@article{VernierPRL24,
  title = {Strong Zero Modes in Integrable Quantum Circuits},
  author = {Vernier, Eric and Yeh, Hsiu-Chung and Piroli, Lorenzo and Mitra, Aditi},
  journal = {Phys. Rev. Lett.},
  volume = {133},
  issue = {5},
  pages = {050606},
  numpages = {7},
  year = {2024},
  month = {Aug},
  publisher = {American Physical Society},
  doi = {10.1103/PhysRevLett.133.050606},
  url = {https://link.aps.org/doi/10.1103/PhysRevLett.133.050606}
}

@article{FendleyPRB14,
  title = {Stability of zero modes in parafermion chains},
  author = {Jermyn, Adam S. and Mong, Roger S. K. and Alicea, Jason and Fendley, Paul},
  journal = {Phys. Rev. B},
  volume = {90},
  issue = {16},
  pages = {165106},
  numpages = {14},
  year = {2014},
  month = {Oct},
  publisher = {American Physical Society},
  doi = {10.1103/PhysRevB.90.165106},
  url = {https://link.aps.org/doi/10.1103/PhysRevB.90.165106}
}

@misc{FendleyXYZ25,
      title={XYZ integrability the easy way}, 
      author={Paul Fendley and Sascha Gehrmann and Eric Vernier and Frank Verstraete},
      year={2025},
      eprint={2511.04674},
      archivePrefix={arXiv},
      primaryClass={cond-mat.stat-mech},
      url={https://arxiv.org/abs/2511.04674}, 
}

@misc{Fendley25,
      title={Strong zero modes in integrable spin-S chains}, 
      author={Fabian H. L. Essler and Paul Fendley and Eric Vernier},
      year={2025},
      eprint={2512.07742},
      archivePrefix={arXiv},
      primaryClass={cond-mat.stat-mech},
      url={https://arxiv.org/abs/2512.07742}, 
}

@article{FendleyJPA16,
doi = {10.1088/1751-8113/49/30/30LT01},
url = {https://doi.org/10.1088/1751-8113/49/30/30LT01},
year = {2016},
month = {jun},
publisher = {IOP Publishing},
volume = {49},
number = {30},
pages = {30LT01},
author = {Fendley, Paul},
title = {Strong zero modes and eigenstate phase transitions in the XYZ/interacting Majorana chain},
journal = {Journal of Physics A: Mathematical and Theoretical}
}

@article{FendleyPRX17,
  title = {Prethermal Strong Zero Modes and Topological Qubits},
  author = {Else, Dominic V. and Fendley, Paul and Kemp, Jack and Nayak, Chetan},
  journal = {Phys. Rev. X},
  volume = {7},
  issue = {4},
  pages = {041062},
  numpages = {22},
  year = {2017},
  month = {Dec},
  publisher = {American Physical Society},
  doi = {10.1103/PhysRevX.7.041062},
  url = {https://link.aps.org/doi/10.1103/PhysRevX.7.041062}
}

@article{review16,
doi = {10.1088/1742-5468/2016/06/064008},
url = {https://doi.org/10.1088/1742-5468/2016/06/064008},
year = {2016},
month = {jun},
publisher = {IOP Publishing and SISSA},
volume = {2016},
number = {6},
pages = {064008},
author = {Ilievski, Enej and Medenjak, Marko and Prosen, Tomaž and Zadnik, Lenart},
title = {Quasilocal charges in integrable lattice systems},
journal = {Journal of Statistical Mechanics: Theory and Experiment}
}

@article{prosen13,
  title = {Families of Quasilocal Conservation Laws and Quantum Spin Transport},
  author = {Prosen, Tomaz and Ilievski, Enej},
  journal = {Phys. Rev. Lett.},
  volume = {111},
  issue = {5},
  pages = {057203},
  numpages = {5},
  year = {2013},
  month = {Aug},
  publisher = {American Physical Society},
  doi = {10.1103/PhysRevLett.111.057203},
  url = {https://link.aps.org/doi/10.1103/PhysRevLett.111.057203}
}

@article{ilievski15,
  title = {Quasilocal Conserved Operators in the Isotropic Heisenberg Spin-$1/2$ Chain},
  author = {Ilievski, Enej and Medenjak, Marko and Prosen, Tomaz},
  journal = {Phys. Rev. Lett.},
  volume = {115},
  issue = {12},
  pages = {120601},
  numpages = {5},
  year = {2015},
  month = {Sep},
  publisher = {American Physical Society},
  doi = {10.1103/PhysRevLett.115.120601},
  url = {https://link.aps.org/doi/10.1103/PhysRevLett.115.120601}
}

@article{GQAI_Nature22,
  title={Formation of robust bound states of interacting microwave photons},
  author={Morvan, A. and Andersen, T.I. and others},
  journal={Nature},
  volume={612},
  number={7939},
  pages={240--245},
  year={2022},
  publisher={Nature Publishing Group UK London},
  url={https://www.nature.com/articles/s41586-022-05348-y}
}

@article{GQAI_Science24a,
  title={Stable quantum-correlated many-body states through engineered dissipation},
  author={Mi, Xiao and Michailidis, Alexios and others},
  journal={Science},
  volume={383},
  pages={1332--1337},
  year={2024},
  publisher={American Association for the Advancement of Science},
  url={
https://doi.org/10.1126/science.adh9932}
}

@article{GQAI_Science24b,
  title={Dynamics of magnetization at infinite temperature in a Heisenberg spin chain},
  author={Rosenberg, E. and Andersen, T.I. and others},
  journal={Science},
  volume={384},
  number={6691},
  pages={48--53},
  year={2024},
  publisher={American Association for the Advancement of Science},
  url={https://doi.org/10.1126/science.adi7877}
}

@article{VanicatPRL18,
  title = {Integrable Trotterization: Local Conservation Laws and Boundary Driving},
  author = {Vanicat, Matthieu and Zadnik, Lenart and Prosen, Tomaz},
  journal = {Phys. Rev. Lett.},
  volume = {121},
  issue = {3},
  pages = {030606},
  numpages = {6},
  year = {2018},
  month = {Jul},
  publisher = {American Physical Society},
  doi = {10.1103/PhysRevLett.121.030606},
  url = {https://link.aps.org/doi/10.1103/PhysRevLett.121.030606}
}

@article{LjubotinaPRL19,
  title = {Ballistic Spin Transport in a Periodically Driven Integrable Quantum System},
  author = {Ljubotina, Marko and Zadnik, Lenart and Prosen, Tomaz},
  journal = {Phys. Rev. Lett.},
  volume = {122},
  issue = {15},
  pages = {150605},
  numpages = {6},
  year = {2019},
  month = {Apr},
  publisher = {American Physical Society},
  doi = {10.1103/PhysRevLett.122.150605},
  url = {https://link.aps.org/doi/10.1103/PhysRevLett.122.150605}
}

@article{Kattel24,
  title = {Kondo effect in the isotropic Heisenberg spin chain},
  author = {Kattel, Pradip and Pasnoori, Parameshwar R. and Pixley, J. H. and Azaria, Patrick and Andrei, Natan},
  journal = {Phys. Rev. B},
  volume = {109},
  issue = {17},
  pages = {174416},
  numpages = {18},
  year = {2024},
  month = {May},
  publisher = {American Physical Society},
  doi = {10.1103/PhysRevB.109.174416},
  url = {https://link.aps.org/doi/10.1103/PhysRevB.109.174416}
}

@misc{Kattel25b,
      title={Thermodynamics in a split Hilbert space: Quantum impurity at the edge of the Heisenberg chain}, 
      author={Abay Zhakenov and Pradip Kattel and Natan Andrei},
      year={2025},
      eprint={2508.19334},
      archivePrefix={arXiv},
      primaryClass={cond-mat.str-el},
      url={https://arxiv.org/abs/2508.19334}, 
}

@article{Kattel25a,
  title = {Quantum Zeno effect in noisy integrable quantum circuits for impurity models},
  author = {Tang, Yicheng and Kattel, Pradip and Pixley, J. H. and Andrei, Natan},
  journal = {Phys. Rev. B},
  volume = {111},
  issue = {5},
  pages = {054313},
  numpages = {15},
  year = {2025},
  month = {Feb},
  publisher = {American Physical Society},
  doi = {10.1103/PhysRevB.111.054313},
  url = {https://link.aps.org/doi/10.1103/PhysRevB.111.054313}
}

@article{KattelXXZ,
  title = {Edge modes and boundary impurities in the anisotropic Heisenberg spin chain},
  author = {Kattel, Pradip and Pasnoori, Parameshwar R. and Pixley, J. H. and Andrei, Natan},
  journal = {Phys. Rev. B},
  volume = {111},
  issue = {17},
  pages = {174430},
  numpages = {36},
  year = {2025},
  month = {May},
  publisher = {American Physical Society},
  doi = {10.1103/PhysRevB.111.174430},
  url = {https://link.aps.org/doi/10.1103/PhysRevB.111.174430}
}

@article{Kemp25,
  title = {Boundary strong zero modes},
  author = {Olund, Christopher T. and Yao, Norman Y. and Kemp, Jack},
  journal = {Phys. Rev. B},
  volume = {111},
  issue = {20},
  pages = {L201114},
  numpages = {6},
  year = {2025},
  month = {May},
  publisher = {American Physical Society},
  doi = {10.1103/PhysRevB.111.L201114},
  url = {https://link.aps.org/doi/10.1103/PhysRevB.111.L201114}
}

@misc{Gehrmann26b,
      title={Exact strong zero modes are generic in integrable spin systems with large anisotropy}, 
      author={Sascha Gehrmann},
      year={2026},
      eprint={2605.26205},
      archivePrefix={arXiv},
      primaryClass={quant-ph},
      url={https://arxiv.org/abs/2605.26205}, 
}

\begin{widetext}
\section*{Appendix: Technical details}
{\small

\noindent
In this Appendix we list some technical details of the exact solution of QLEM, in particular lengthy analytical expressions of various auxiliary and observable quantities. These objects are all
referenced from the main text. 
\\\\
Pauli basis representation of a 2-qubit SU(2)-symmetric six vertex gate reads:
\be
W(\tau)=
\frac{1}{1+\tau^2}
\left(
\begin{array}{cccccccccccccccc}
 \tau ^2+1 & 0 & 0 & 0 & 0 & 0 & 0 & 0 & 0 & 0 & 0 & 0 & 0 & 0 & 0 & 0 \\
 0 & 1 & 0 & 0 & \tau ^2 & 0 & 0 & 0 & 0 & 0 & 0 & -\tau  & 0 & 0 & \tau  & 0 \\
 0 & 0 & 1 & 0 & 0 & 0 & 0 & \tau  & \tau ^2 & 0 & 0 & 0 & 0 & -\tau  & 0 & 0 \\
 0 & 0 & 0 & 1 & 0 & 0 & -\tau  & 0 & 0 & \tau  & 0 & 0 & \tau ^2 & 0 & 0 & 0 \\
 0 & \tau ^2 & 0 & 0 & 1 & 0 & 0 & 0 & 0 & 0 & 0 & \tau  & 0 & 0 & -\tau  & 0 \\
 0 & 0 & 0 & 0 & 0 & \tau ^2+1 & 0 & 0 & 0 & 0 & 0 & 0 & 0 & 0 & 0 & 0 \\
 0 & 0 & 0 & \tau  & 0 & 0 & 1 & 0 & 0 & \tau ^2 & 0 & 0 & -\tau  & 0 & 0 & 0 \\
 0 & 0 & -\tau  & 0 & 0 & 0 & 0 & 1 & \tau  & 0 & 0 & 0 & 0 & \tau ^2 & 0 & 0 \\
 0 & 0 & \tau ^2 & 0 & 0 & 0 & 0 & -\tau  & 1 & 0 & 0 & 0 & 0 & \tau  & 0 & 0 \\
 0 & 0 & 0 & -\tau  & 0 & 0 & \tau ^2 & 0 & 0 & 1 & 0 & 0 & \tau  & 0 & 0 & 0 \\
 0 & 0 & 0 & 0 & 0 & 0 & 0 & 0 & 0 & 0 & \tau ^2+1 & 0 & 0 & 0 & 0 & 0 \\
 0 & \tau  & 0 & 0 & -\tau  & 0 & 0 & 0 & 0 & 0 & 0 & 1 & 0 & 0 & \tau ^2 & 0 \\
 0 & 0 & 0 & \tau ^2 & 0 & 0 & \tau  & 0 & 0 & -\tau  & 0 & 0 & 1 & 0 & 0 & 0 \\
 0 & 0 & \tau  & 0 & 0 & 0 & 0 & \tau ^2 & -\tau  & 0 & 0 & 0 & 0 & 1 & 0 & 0 \\
 0 & -\tau  & 0 & 0 & \tau  & 0 & 0 & 0 & 0 & 0 & 0 & \tau ^2 & 0 & 0 & 1 & 0 \\
 0 & 0 & 0 & 0 & 0 & 0 & 0 & 0 & 0 & 0 & 0 & 0 & 0 & 0 & 0 & \tau ^2+1 \\
\end{array}
\right)
\label{eq:W}
\ee
Here we list explicit representation of the boundary vectors and bulk matrices of MPA representation of QLEM, specifically the boundary vectors (note that $r= \sqrt{\omega ^3 \left(\left(4-\tau ^2\right) \omega -4 \tau \right)}$):
\bea
\bra{a^0}&=& 
\left(
\begin{array}{cccccccccccccccc}
 0 & 0 & 0 & 0 & 0 & 0 & 0 & 0 & 0 & 0 & 0 & 0 & 0 & 0 & 0 & \frac{\left(\tau ^2+5\right) \omega ^3 +\tau  \omega ^2 +\tau r}{\omega  \left(\left(\tau ^2+5\right) \omega ^2+\tau ^2-2 \tau  \omega \right)} \\
\end{array}
\right)\,, \nonumber\\
\bra{a^1}&=&
\left(
\begin{array}{cccccccccccccccc}
 1 & \frac{2 (\tau -2 \omega ) \left(3 \tau  \omega ^2+\tau +5 \omega
   \right)+2  (2 \tau  \omega +5)r}{\tau  \omega  \left(4 \tau  \omega ^2+\tau +8 \omega \right)} & 0 & 0 & 0 & 0 & 0 & 0 & 0 & 0 & 0 & 0 & 0 & 0 & 0 & 0 \\
\end{array}
\right)\,, \label{eq:a}\\
\bra{a^2}&=& \bra{a^1}P\,,\qquad \bra{a^3}= \bra{a^2}P\,,\qquad \bra{a^1}= \bra{a^3}P\,, \nonumber
\eea
and the bulk matrices:
\be
 A^0=
 \left(
\begin{array}{cccccccccccccccc}
 1 & 0 & 0 & 0 & 0 & 0 & 0 & 0 & 0 & 0 & 0 & 0 & 0 & 0 & 0 & 0 \\
 0 &\!\!x_1\!+\!x_2\!+\!x_3\!\!& 0 & 0 & 0 & 0 & 0 & 0 & 0 & x_4 & 0 & 0 & 0 & 0 & 0 & 0 \\
 0 & 0 & 1 & 0 & 0 & 0 & 0 & 0 & 0 & 0 & 0 & 0 & 0 & 0 & 0 & 0 \\
 0 & 0 & 0 &\!\!x_1\!+\!x_2\!+\!x_3\!\!& 0 & 0 & 0 & 0 & 0 & 0 & 0 & x_4 & 0 & 0 & 0 & 0 \\
 0 & 0 & 0 & 0 & 1 & 0 & 0 & 0 & 0 & 0 & 0 & 0 & 0 & 0 & 0 & 0 \\
 0 & 0 & 0 & 0 & 0 &\!\!x_1\!+\!x_2\!+\!x_3\!\!& 0 & x_4 & 0 & 0 & 0 & 0 & 0 & 0 & 0 & 0 \\
 0 & 0 & 0 & 0 & 0 & 0 & -\frac{x_2}{2} & 0 & 0 & 0 & 0 & 0 & 0 & 0 & 0 & 0 \\
 0 & 0 & 0 & 0 & 0 & x_1 & 0 &\!\!x_1\!+\!x_2\!+\!x_3\!\!& 0 & 0 & 0 & 0 & 0 & 0 & 0 & 0 \\
 0 & 0 & 0 & 0 & 0 & 0 & 0 & 0 & -\frac{x_2}{2} & 0 & 0 & 0 & 0 & 0 & 0 & 0 \\
 0 & x_1 & 0 & 0 & 0 & 0 & 0 & 0 & 0 &\!\!x_1\!+\!x_2\!+\!x_3\!\!& 0 & 0 & 0 & 0 & 0 & 0 \\
 0 & 0 & 0 & 0 & 0 & 0 & 0 & 0 & 0 & 0 & -\frac{x_2}{2} & 0 & 0 & 0 & 0 & 0 \\
 0 & 0 & 0 & x_1 & 0 & 0 & 0 & 0 & 0 & 0 & 0 &\!\!x_1\!+\!x_2\!+\!x_3\!\!& 0 & 0 & 0 & 0 \\
 0 & 0 & 0 & 0 & 0 & 0 & 0 & 0 & 0 & 0 & 0 & 0 & \frac{x_3}{3} & \frac{x_2}{2}\!+\!\frac{x_3}{3} & \!\frac{x_2}{2}\!+\!\frac{x_3}{3}\! & 0 \\
 0 & 0 & 0 & 0 & 0 & 0 & 0 & 0 & 0 & 0 & 0 & 0 & \!\frac{x_2}{2}\!+\!\frac{x_3}{3}\! & \frac{x_3}{3} & \!\frac{x_2}{2}\!+\!\frac{x_3}{3}\! & 0 \\
 0 & 0 & 0 & 0 & 0 & 0 & 0 & 0 & 0 & 0 & 0 & 0 & \!\frac{x_2}{2}\!+\!\frac{x_3}{3}\! & \!\frac{x_2}{2}\!+\!\frac{x_3}{3}\! & \frac{x_3}{3} & 0 \\
 0 & 0 & 0 & 0 & 0 & 0 & 0 & 0 & 0 & 0 & 0 & 0 & 0 & 0 & 0 & x_5 \\
\end{array}
\right)
 \nonumber
 \ee
 \bea
 A^1&=&
 \left(
\begin{array}{cccccccccccccccc}
 0 & 0 & 0 & 0 & 0 & 0 & 0 & 0 & 0 & 0 & 0 & 0 & 2 x_6\!+\!x_7 & x_7\!-\!x_6 & x_7\!-\!x_6 & x_{11} \\
 0 & 0 & 0 & 0 & 0 & 0 & 0 & 0 & 0 & 0 & 0 & 0 & 1 & 1 & 1 & 0 \\
 0 & 0 & 0 & 0 & -\frac{2 x_8}{3} & -x_9 & \frac{2 x_{10}}{3} & x_{10} & 0 & 0 & 0 & 0 & 0 & 0 & 0 & 0 \\
 0 & 0 & 0 & 0 & 0 & -x_{12} & 0 & -x_{13} & 0 & 0 & 0 & 0 & 0 & 0 & 0 & 0 \\
 0 & 0 & \frac{2 x_8}{3} & x_9 & 0 & 0 & 0 & 0 & 0 & 0 & \frac{2 x_{10}}{3} & -x_{10} & 0 & 0 & 0 & 0 \\
 0 & 0 & 0 & x_{12} & 0 & 0 & 0 & 0 & 0 & 0 & 0 & x_{13} & 0 & 0 & 0 & 0 \\
 0 & 0 & x_{14} & x_{15} & 0 & 0 & 0 & 0 & 0 & 0 & -\frac{2 x_8}{3} & -x_8 & 0 & 0 & 0 & 0 \\
 0 & 0 & 0 & -x_{16} & 0 & 0 & 0 & 0 & 0 & 0 & 0 & x_{12} & 0 & 0 & 0 & 0 \\
 0 & 0 & 0 & 0 & 0 & 0 & 0 & 0 & 0 & 0 & 0 & 0 & 0 & -x_{17} & x_{17} & 0 \\
 0 & 0 & 0 & 0 & 0 & 0 & 0 & 0 & 0 & 0 & 0 & 0 & 1 & 1 & 1 & 0 \\
 0 & 0 & 0 & 0 & x_{14} & x_{15} & \frac{2 x_8}{3} & -x_8 & 0 & 0 & 0 & 0 & 0 & 0 & 0 & 0 \\
 0 & 0 & 0 & 0 & 0 & x_{16} & 0 & -x_{12} & 0 & 0 & 0 & 0 & 0 & 0 & 0 & 0 \\
 -2 x_{18} & 2 x_{19} & 0 & 0 & 0 & 0 & 0 & 0 & 0 & 1 & 0 & 0 & 0 & 0 & 0 & 0 \\
 x_{18} & -x_{19} & 0 & 0 & 0 & 0 & 0 & 0 & 1 & -\frac{1}{2} & 0 & 0 & 0 & 0 & 0 & 0 \\
 x_{18} & -x_{19} & 0 & 0 & 0 & 0 & 0 & 0 & -1 & -\frac{1}{2} & 0 & 0 & 0 & 0 & 0 & 0 \\
 1 & x_{20} & 0 & 0 & 0 & 0 & 0 & 0 & 0 & x_{21} & 0 & 0 & 0 & 0 & 0 & 0 \\
\end{array}
\right) \label{eq:AA}\\
A^2&=& P^{-1} A^1 P\,,\qquad A^3= P^{-1} A^2 P\,,\qquad A^1= P^{-1} A^3 P\,, \nonumber
 \end{eqnarray}
where $x_i=u_i + v_i r$ with
\begin{eqnarray*}
u_{1}&=&\frac{\omega  \left(-2 \tau ^3+\left(7 \tau ^2+18\right) \tau  \omega ^2-\left(\tau ^2-4\right) \tau ^2 \omega +\left(\tau ^4-2 \tau ^2-20\right) \omega ^3\right)}{2 \tau ^2 \left(\omega ^2+1\right) \left(\left(\tau ^2+5\right) \omega ^2+\tau ^2-2 \tau  \omega \right)},\quad
u_{2}=\frac{2 \omega ^2-2 \tau ^2 \left(\omega ^2+1\right)}{\left(\tau ^2+5\right) \omega ^2+\tau ^2-2 \tau  \omega },\cr
u_{3}&=&\frac{\tau ^4 \left(2 \omega ^4+7 \omega ^2+3\right)+\tau ^3 \left(\omega -8 \omega ^3\right)+\tau ^2 \omega ^2 \left(3 \omega ^2-5\right)-9 \tau  \omega ^3+10 \omega ^4}{\tau ^2 \left(\omega ^2+1\right) \left(\left(\tau ^2+5\right) \omega ^2+\tau ^2-2 \tau  \omega \right)},\cr
u_{4}&=&\frac{\tau ^2 \omega ^2 \left(-6 \tau ^3 \omega +\tau  \left(11 \tau ^2-30\right) \omega ^3-8 \tau ^2+\left(\tau ^4-2 \tau ^2+4\right) \omega ^4-\left(\tau ^4-48 \tau ^2+56\right) \omega ^2+64 \tau  \omega \right)}{2 \left(\omega ^2+1\right) \left(\left(\tau ^2+5\right) \omega ^2+\tau ^2-2 \tau  \omega \right) \left(\left(\tau ^4-12\right) \omega ^2+4 \left(\tau ^2+2\right) \tau  \omega +4 \tau ^2\right)},\quad
u_{5}=\frac{\left(\tau ^2+8\right) \omega ^2+\tau ^2-3 \tau  \omega }{\left(\tau ^2+5\right) \omega ^2+\tau ^2-2 \tau  \omega },\cr
u_{6}&=&\frac{2 \tau  \omega ^2 \left(-82 \tau ^4 \omega -88 \tau ^3+240 \tau ^2 \omega +4 \left(\tau ^4-4 \tau ^2-6\right) \tau  \omega ^4+\left(-27 \tau ^4+188 \tau ^2-104\right) \tau  \omega ^2-3 \left(\tau ^6-17 \tau ^4+22 \tau ^2+16\right) \omega ^3\right)}{9 \left(4 \tau  \omega ^2+\tau +8 \omega \right) \left(\left(\tau ^2+5\right) \omega ^2+\tau ^2-2 \tau  \omega \right) \left(\left(\tau ^4-12\right) \omega ^2+4 \left(\tau ^2+2\right) \tau  \omega +4 \tau ^2\right)},\cr
u_{7}&=&\frac{4 \left(2 \tau\!\left(\tau ^2\!\!+\!5\right) \left(\tau ^4\!\!-\!\!\tau ^2\!\!-\!\!14\right) \omega ^5\!\!-\!18 \tau ^4\!+\!\tau\!\left(58 \tau ^4\!+\!199 \tau ^2\!\!-\!42\right) \omega ^3\!\!-\!2 \tau ^2\!\left(\tau ^4\!\!-\!47 \tau ^2\!\!-\!\!111\right) \omega ^2\!\!+\!\tau ^3\!\left(58\!-\!11 \tau ^2\right) \omega\!+\!\left(17 \tau ^6\!\!+\!77 \tau ^4\!\!-\!34 \tau ^2\!-\!\!220\right) \omega ^4\right)}{\tau  \left(\left(\tau ^2+5\right) \omega +4 \tau \right) \left(4 \tau  \omega ^2+\tau +8 \omega \right) \left(\left(\tau ^4-12\right) \omega ^2+4 \left(\tau ^2+2\right) \tau  \omega +4 \tau ^2\right)},\cr
u_{8}&=&-\frac{3 \tau  \omega  (\tau -3 \omega )}{4 \left(\left(\tau ^2+5\right) \omega ^2+\tau ^2-2 \tau  \omega \right)},\quad
u_{9}=\frac{\omega  (4 \omega -\tau  (2 \tau  \omega +7))}{\tau  \left(4 \tau  \omega ^2+\tau +8 \omega \right)},\\
u_{10}&=&\frac{\omega  \left(-50 \tau ^3 \omega -56 \tau ^2-2 \left(\tau ^4-2 \tau ^2-12\right) \tau  \omega ^3+\left(-17 \tau ^4+6 \tau ^2+48\right) \omega ^2+8 \tau  \omega \right)}{\left(4 \tau  \omega ^2+\tau +8 \omega \right) \left(\left(\tau ^4-12\right) \omega ^2+4 \left(\tau ^2+2\right) \tau  \omega +4 \tau ^2\right)},\cr
u_{11}&=&\frac{2 \omega  \left(-4 \tau ^3+6 \left(\tau ^2+1\right) \tau  \omega ^2-\left(\tau ^2-15\right) \tau ^2 \omega +\left(\tau ^4+4 \tau ^2-5\right) \omega ^3\right)}{3 \left(\left(\tau ^2+5\right) \omega ^2+\tau ^2-2 \tau  \omega \right)^2},\quad
u_{12}=-\frac{\omega  (\tau -2 \omega )}{2 \tau  \left(\omega ^2+1\right)},\cr
u_{13}&=&\frac{\tau  \omega  \left(-6 \tau ^4 \omega -8 \tau ^3+8 \tau ^2 \omega +2 \left(\tau ^4-4 \tau ^2-6\right) \omega ^5-\tau  \left(\tau ^4-23 \tau ^2+22\right) \omega ^4-2 \left(3 \tau ^4-29 \tau ^2+36\right) \omega ^3-\tau  \left(\tau ^4+2 \tau ^2-72\right) \omega ^2\right)}{2 \left(\omega ^2+1\right) \left(\left(\tau ^2+5\right) \omega ^2+\tau ^2-2 \tau  \omega \right) \left(\left(\tau ^4-12\right) \omega ^2+4 \left(\tau ^2+2\right) \tau  \omega +4 \tau ^2\right)},\cr
u_{14}&=&\Bigl(3 \omega  \left(\left(\tau ^2\!-\!4\right) \omega +4 \tau \right) \bigl(24 \tau ^4-2 \tau  \left(46 \tau ^4\!+\!249 \tau ^2\!+\!116\right) \omega ^3+\tau ^2 \left(\tau ^4\!-\!178 \tau ^2\!-\!456\right) \omega ^2-2 \tau ^3 \left(76\!-\!5 \tau ^2\right) \omega\cr
&-& 2 \tau  \left(\tau ^6\!+\!7 \tau ^4\!-\!74 \tau ^2\!-\!300\right) \omega ^5-\left(23 \tau ^6\!+\!185 \tau ^4\!-\!114 \tau ^2\!-\!1200\right) \omega ^4\bigr)\Bigr)/\Bigl(8 \tau  (\tau  \omega +4)^2 \left(\left(\tau ^2-3\right) \omega +4 \tau \right) \left(\left(\tau ^2+5\right) \omega ^2+\tau ^2-2 \tau  \omega \right)^2\Bigr),\cr
u_{15}&=&\frac{3 \omega  \left(-8 \tau ^3+2 \left(8-3 \tau ^2\right) \tau ^2 \omega +\left(5 \tau ^4-6 \tau ^2-80\right) \omega ^3+\left(-\tau ^4+28 \tau ^2+72\right) \tau  \omega ^2\right)}{4 \tau  (\tau  \omega +4)^2 \left(\left(\tau ^2+5\right) \omega ^2+\tau ^2-2 \tau  \omega \right)},\cr
u_{16}&=&\frac{-2 \left(\tau ^4-10\right) \omega ^4+2 \tau ^4+(\tau -3) (\tau +3) \left(\tau ^2+2\right) \tau  \omega ^3+2 \left(\tau ^2-3\right) \tau ^2 \omega ^2+\left(\tau ^2+2\right) \tau ^3 \omega }{2 \tau ^3 \left(\omega ^2+1\right) \left(\left(\tau ^2+5\right) \omega ^2+\tau ^2-2 \tau  \omega \right)},\quad
u_{17}=\frac{\tau ^2 \omega ^2 \left(\tau ^2-\left(\tau ^2-13\right) \omega ^2-10 \tau  \omega \right)}{2 \left(\left(\tau ^2+5\right) \omega ^2+\tau ^2-2 \tau  \omega \right)^2},\cr
u_{18}&=&-\frac{\left(\left(\tau ^2-4\right) \omega +4 \tau \right) \left(56 \tau ^3-8 \tau  \left(4 \tau ^2+15\right) \omega ^4+2 \tau ^2 \left(21 \tau ^2-8\right) \omega +\left(\tau ^2+8\right) \left(\tau ^4-3 \tau ^2-30\right) \omega ^3+\tau  \left(11 \tau ^4+4 \tau ^2+8\right) \omega ^2\right)}{4 \tau ^2 (\tau  \omega +4)^2 \left(\left(\tau ^2-3\right) \omega +4 \tau \right) \left(\left(\tau ^2+5\right) \omega ^2+\tau ^2-2 \tau  \omega \right)},\cr
u_{19}&=&\frac{-2 \left(3 \tau ^2+4\right) \tau  \omega -8 \tau ^2+\left(-\tau ^4+2 \tau ^2+16\right) \omega ^2}{2 \tau ^2 (\tau  \omega +4)^2},\quad
u_{20}=\frac{-8 \tau  \left(\tau ^2+5\right) \omega ^3+\tau  \left(7 \tau ^2-66\right) \omega +18 \tau ^2+\left(\tau ^4-33 \tau ^2-60\right) \omega ^2}{\tau  \left(\left(\tau ^2+5\right) \omega +4 \tau \right) \left(4 \tau  \omega ^2+\tau +8 \omega \right)},\cr
u_{21}&=&\frac{\tau  \omega  \left(-56 \tau ^3+8 \tau  \left(\tau ^2+5\right) \omega ^4+6 \tau ^2 \left(8-7 \tau ^2\right) \omega +\tau  \left(-11 \tau ^4+12 \tau ^2-72\right) \omega ^2-\left(\tau ^6+\tau ^4+10 \tau ^2-80\right) \omega ^3\right)}{\left(\left(\tau ^2+5\right) \omega +4 \tau \right) \left(4 \tau  \omega ^2+\tau +8 \omega \right) \left(\left(\tau ^4-12\right) \omega ^2+4 \left(\tau ^2+2\right) \tau  \omega +4 \tau ^2\right)},\cr
 \end{eqnarray*}
\begin{eqnarray*}
 v_{1}&=&\frac{1}{2} \left(\frac{\tau ^2+4}{\left(\tau ^2+5\right) \omega ^2+\tau ^2-2 \tau  \omega }-\frac{\tau ^2+2}{\tau ^2 \left(\omega ^2+1\right)}\right),\quad
v_{2}=\frac{4}{\left(\tau ^2+5\right) \omega ^2+\tau ^2-2 \tau  \omega },\quad
v_{3}=-\frac{(\tau +\omega ) (\tau  (2 \tau  \omega +7)-5 \omega )}{\tau ^2 \left(\omega ^2+1\right) \left(\left(\tau ^2+5\right) \omega ^2+\tau ^2-2 \tau  \omega \right)},\cr
v_{4}&=&\frac{1}{2} \tau ^2 \left(\frac{1}{\left(\tau ^2+5\right) \omega ^2+\tau ^2-2 \tau  \omega }-\frac{1}{\left(\tau ^2+2\right) \left(\omega ^2+1\right)}-\frac{8}{\left(\tau ^2+2\right) \left(\left(\tau ^4-12\right) \omega ^2+4 \left(\tau ^2+2\right) \tau  \omega +4 \tau ^2\right)}\right),\cr
v_{5}&=&\frac{1}{\left(\tau ^2+5\right) \omega ^2+\tau ^2-2 \tau  \omega },\quad
v_{6}=\frac{2 \tau  \left(-6 \tau ^4 \omega -8 \tau ^3+88 \tau ^2 \omega +2 \left(\tau ^4+\tau ^2+6\right) \tau  \omega ^4+6 \left(3 \tau ^4-3 \tau ^2+4\right) \omega ^3-\left(\tau ^4-68 \tau ^2+56\right) \tau  \omega ^2\right)}{9 \left(4 \tau  \omega ^2+\tau +8 \omega \right) \left(\left(\tau ^2+5\right) \omega ^2+\tau ^2-2 \tau  \omega \right) \left(\left(\tau ^4-12\right) \omega ^2+4 \left(\tau ^2+2\right) \tau  \omega +4 \tau ^2\right)},\cr
v_{7}&=&\frac{8 \tau  \left(\tau ^2\!+\!4\right) \left(\tau ^2\!+\!5\right) \omega ^5  -24 \tau ^4-   4 \tau  \left(18 \tau ^4\!+\!35 \tau ^2\!-\!46\right) \omega ^3-4 \tau ^2 \left(\tau ^4\!+\!35 \tau ^2\!+\!52\right) \omega ^2-4 \tau ^3 \left(5 \tau ^2\!+\!26\right) \omega -4 \left(3 \tau ^6\!-\!2 \tau ^4\!-\!73 \tau ^2\!-\!110\right) \omega ^4}{\tau  \omega ^2 \left(\left(\tau ^2+5\right) \omega +4 \tau \right) \left(4 \tau  \omega ^2+\tau +8 \omega \right) \left(\left(\tau ^4-12\right) \omega ^2+4 \left(\tau ^2+2\right) \tau  \omega +4 \tau ^2\right)},\cr
v_{8}&=&\frac{3 \tau }{4 \left(\left(\tau ^2+5\right) \omega ^2+\tau ^2-2 \tau  \omega \right)},\quad
v_{9}=\frac{2 \omega -\tau }{\tau  \omega  \left(4 \tau  \omega ^2+\tau +8 \omega \right)},\quad
v_{10}=-\frac{(\tau  \omega +2) \left(2 \left(\tau ^2+6\right) \omega ^2+\left(\tau ^2-4\right) \tau  \omega +4 \tau ^2\right)}{\omega  \left(4 \tau  \omega ^2+\tau +8 \omega \right) \left(\left(\tau ^4-12\right) \omega ^2+4 \left(\tau ^2+2\right) \tau  \omega +4 \tau ^2\right)},\cr
v_{11}&=&\frac{-4 \left(\tau ^2+5\right) \omega ^2+4 \left(\tau ^2+1\right) \tau  \omega +8 \tau ^2}{3 \left(\left(\tau ^2+5\right) \omega ^2+\tau ^2-2 \tau  \omega \right)^2},\quad
v_{12}=\frac{1}{2 \tau  \omega ^2+2 \tau },\cr
v_{13}&=&\tau  \left(-\frac{1}{2 \left(\tau ^2+2\right) \left(\omega ^2+1\right)}+\frac{1}{\left(\tau ^2+5\right) \omega ^2+\tau ^2-2 \tau  \omega }+\frac{2 \tau ^2}{\left(\tau ^2+2\right) \left(\left(\tau ^4-12\right) \omega ^2+4 \left(\tau ^2+2\right) \tau  \omega +4 \tau ^2\right)}\right),\cr
v_{14}&=&
\Bigl(-96 \tau ^5-3 \tau ^2 \left(\tau ^2+4\right) \left(\tau ^4+116 \tau ^2-368\right) \omega ^3-96 \tau ^4 \left(\tau ^2+3\right) \omega +6 \tau  \left(3 \tau ^6+37 \tau ^4-82 \tau ^2-600\right) \omega ^6 \cr
&-&6 \tau  \left(16 \tau ^6-37 \tau ^4-796 \tau ^2-688\right) \omega ^4-6 \tau ^3 \left(5 \tau ^4+92 \tau ^2+160\right) \omega ^2+3 \left(-3 \tau ^8+45 \tau ^6+668 \tau ^4+528 \tau ^2-2400\right) \omega ^5\Bigr)/\cr
&/&\Bigl(
8 \tau  \omega  (\tau  \omega +4)^2 \left(\left(\tau ^2-3\right) \omega +4 \tau \right) \left(\left(\tau ^2+5\right) \omega ^2+\tau ^2-2 \tau  \omega \right)^2\Bigr),\cr
v_{15}&=&\frac{3 \tau }{4 \left(\left(\tau ^2+5\right) \omega ^2+\tau ^2-2 \tau  \omega \right)}-\frac{6}{\tau  (\tau  \omega +4)^2},\quad
v_{16}=\frac{\tau ^2+2}{2 \tau ^3 \left(\omega ^2+1\right)}-\frac{\tau ^2+4}{\tau  \left(\left(\tau ^2+5\right) \omega ^2+\tau ^2-2 \tau  \omega \right)},\quad
v_{17}=-\frac{\tau ^2 \omega  (\tau -3 \omega )}{\left(\left(\tau ^2+5\right) \omega ^2+\tau ^2-2 \tau  \omega \right)^2},\cr
v_{18}&=&\frac{\frac{8 \left(\tau ^2+8\right) \tau }{\left(\tau ^2+5\right) \omega ^2+\tau ^2-2 \tau  \omega }-\frac{4 \left(\tau ^2+4\right) \tau }{\tau  \omega +4}+\frac{\tau ^2+7}{\omega }+\frac{3 \left(\tau ^4-9\right)}{\left(\tau ^2-3\right) \omega +4 \tau }+\frac{4 \tau }{\omega ^2}+\frac{64 \tau }{(\tau  \omega +4)^2}}{32 \tau ^2},\quad v_{19}=\frac{\tau ^2+4}{\tau ^2 (\tau  \omega +4)^2},\cr
v_{20}&=&\frac{2 \tau  \left(\tau ^2+5\right) \omega ^3+2 \left(8 \tau ^2+15\right) \omega ^2+\tau  \left(\tau ^2+18\right) \omega +6 \tau ^2}{\tau  \omega ^2 \left(\left(\tau ^2+5\right) \omega +4 \tau \right) \left(4 \tau  \omega ^2+\tau +8 \omega \right)},\cr
v_{21}&=&\frac{\tau  \left(-6 \tau ^4 \omega -8 \tau ^3-2 \left(\tau ^2+2\right) \left(\tau ^2+5\right) \tau  \omega ^4+8 \tau ^2 \omega -2 \left(6 \tau ^4+35 \tau ^2+20\right) \omega ^3-\left(\tau ^4+16 \tau ^2+104\right) \tau  \omega ^2\right)}{\omega  \left(\left(\tau ^2+5\right) \omega +4 \tau \right) \left(4 \tau  \omega ^2+\tau +8 \omega \right) \left(\left(\tau ^4-12\right) \omega ^2+4 \left(\tau ^2+2\right) \tau  \omega +4 \tau ^2\right)}.
\end{eqnarray*}
The leading right eigenvector of the transfer matrix $\mathbb T$ reads (writing $\ket{i,j}=\ket{i}\otimes\ket{j}$):
\bea
\ket{\mathrm{R}} &=& z_1 (\ket{1,1}+\ket{3,3}+\ket{5,5}) + z_2  (\ket{2,2}+\ket{4,4}+\ket{6,6}) 
+\ket{1,2}+\ket{2,1}
+\ket{3,4}+\ket{4,3}
+\ket{5,6}+\ket{6,5} \nonumber\\
&+&z_3 (\ket{7,7}+\ket{9,9}+\ket{11,11}) + z_4  (\ket{8,8}+\ket{10,10}+\ket{12,12}) \label{RR}\\
&+& z_5 (\ket{13,13}+\ket{14,14}+\ket{15,15})+z_6(\ket{13,14}+\ket{13,15}+\ket{14,15}+\ket{14,13}+\ket{15,13}+\ket{15,14})+z_7 \ket{16,16}\,. \nonumber
\eea
where $z_i=p_i + q_i r$ with
\begin{eqnarray*}
p_{1}&=&\frac{4 \tau ^3+2 \tau  \left(\tau ^2-6\right) \left(\tau ^2+5\right) \omega ^4+3 \tau  \left(29 \tau ^2+2\right) \omega ^2+\tau ^2 \left(\tau ^2+94\right) \omega +2 \left(12 \tau ^4-\tau ^2-70\right) \omega ^3}{\tau  \omega  \left(\left(\tau ^2+5\right) \omega +4 \tau \right) \left(4 \tau  \omega ^2+\tau +8 \omega \right)},\cr
p_{2}&=&\frac{\omega  \left(-12 \tau ^3+3 \tau  \left(9 \tau ^2+10\right) \omega ^2-3 \tau ^2 \left(\tau ^2-18\right) \omega +2 \tau  \left(\tau ^4+3 \tau ^2-22\right) \omega ^4+6 \left(2 \tau ^4+9 \tau ^2-10\right) \omega ^3\right)}{4 \left(\omega ^2+1\right) \left(\left(\tau ^2-3\right) \omega +4 \tau \right) \left(\left(\tau ^2+5\right) \omega ^2+\tau ^2-2 \tau  \omega \right)},\cr
p_{3}&=&\Big(9 \Bigl(-176 \tau ^6+4 \left(109-55 \tau ^2\right) \tau ^5 \omega -2 \left(\tau ^2 \left(\tau ^2-3\right) \left(\tau ^2+5\right) \left(\tau ^4-12\right)-288\right) \omega ^6-4 \left(26 \tau ^4-66 \tau ^2+259\right) \tau ^4 \omega ^2\cr
&-&2 \left(10 \tau ^8+35 \tau ^6-143 \tau ^4-264 \tau ^2+924\right) \tau  \omega ^5+\left(-2 \tau ^8-67 \tau ^6-341 \tau ^4+636 \tau ^2+1564\right) \tau ^2 \omega ^4-\left(23 \tau ^6+43 \tau ^4+920 \tau ^2-484\right) \tau ^3 \omega ^3\Bigr)\Bigr)/\cr
&/&\Bigl(16 \tau  \omega  \left(\left(\tau ^2-3\right) \omega +4 \tau \right)^2 \left(\left(\tau ^2+5\right) \omega +4 \tau \right) \left(\left(\tau ^2+5\right) \omega ^2+\tau ^2-2 \tau  \omega \right)\Bigr),\cr
p_{4}&=&\frac{\tau  \left(11 \tau ^4\!+\!136 \tau ^2\!+\!124\right) \omega ^3\!-28 \tau ^4\!-\tau ^2 \left(3 \tau ^4\!-\!4 \tau ^2\!+\!12\right) \omega ^2\!+4 \tau ^3 \left(9\!-\!5 \tau ^2\right) \omega +2 \tau  \left(\tau ^6\!+\!\tau ^4\!-\!20 \tau ^2\!-\!12\right) \omega ^5\!+2 \left(6 \tau ^6\!+\!31 \tau ^4\!-\!36 \tau ^2\!-\!60\right) \omega ^4}{4 \tau ^2 \left(\omega ^2+1\right) \left(\left(\tau ^2-3\right) \omega +4 \tau \right) \left(\left(\tau ^2+5\right) \omega ^2+\tau ^2-2 \tau  \omega \right)},\cr
p_{5}&=&\Bigl(-16416 \tau ^{13}+216 \tau ^{12} \left(1293-103 \tau ^2\right) \omega -216 \tau ^{11} \left(52 \tau ^4-1194 \tau ^2+8965\right) \omega ^2-54 \tau ^{10} \left(47 \tau ^6-757 \tau ^4+14396 \tau ^2-157832\right) \omega ^3\cr
&-&54 \tau ^8 \left(208 \tau ^8-7411 \tau ^6+66461 \tau ^4-181320 \tau ^2-974888\right) \omega ^5\cr
&+&2 \tau ^6 \left(-8019 \tau ^{10}+354942 \tau ^8-839864 \tau ^6-10674697 \tau ^4+51362208 \tau ^2+19364400\right) \omega ^7 \cr
&-&54 \tau ^9 \left(4 \tau ^8+507 \tau ^6-12093 \tau ^4-3668 \tau ^2+470560\right) \omega ^4 \cr
&+&2 \tau  \left(\tau ^2+5\right)^2 \left(55 \tau ^{12}-613 \tau ^{10}-2835 \tau ^8+25897 \tau ^6+17732 \tau ^4-284700 \tau ^2+396000\right) \omega ^{18} \cr
&+&2 \tau ^4 \left(-270 \tau ^{12}\!+\!224994 \tau ^{10}\!+\!2202476 \tau ^8\!-\!16631051 \tau ^6\!+\!2539633 \tau ^4\!+\!25017876 \tau ^2\!-\!81649620\right) \omega ^9 \cr
&-&2 \tau ^7 \left(594 \tau ^{10}\!-\!16308 \tau ^8\!+\!96579 \tau ^6\!-\!6159583 \tau ^4\!+\!21565656 \tau ^2\!+\!35163288\right) \omega ^6 \cr
&+&2 \left(\tau ^2+5\right) \left(1075 \tau ^{14}+2942 \tau ^{12}-78808 \tau ^{10}-235858 \tau ^8+1377145 \tau ^6+2507620 \tau ^4-9476100 \tau ^2+72000\right) \omega ^{17} \cr
&+&\tau ^2 \left(11151 \tau ^{14}+38816 \tau ^{12}-444304 \tau ^{10}+470818 \tau ^8+12518757 \tau ^6-9771242 \tau ^4-138044980 \tau ^2-279682600\right) \omega ^{15}\cr 
&+&\tau ^2 \left(25464 \tau ^{14}-72821 \tau ^{12}+905837 \tau ^{10}+11220545 \tau ^8-14775797 \tau ^6-107979648 \tau ^4-471479444 \tau ^2-280422000\right) \omega ^{13} \cr
&+&\tau ^2 \left(22483 \tau ^{14}+29653 \tau ^{12}+4541935 \tau ^{10}-2687213 \tau ^8-72353574 \tau ^6-62273060 \tau ^4-482964120 \tau ^2-93474000\right) \omega ^{11} \cr
&-&2 \tau ^5 \left(1296 \tau ^{12}-92205 \tau ^{10}+731506 \tau ^8-6790254 \tau ^6-9452717 \tau ^4+69986592 \tau ^2-28888272\right) \omega ^8 \cr
&+&\tau  \left(408 \tau ^{16}+5923 \tau ^{14}+109068 \tau ^{12}+431366 \tau ^{10}-3849636 \tau ^8-21109581 \tau ^6+17630680 \tau ^4+157394300 \tau ^2+56112000\right) \omega ^{16}\cr
&+&\tau  \left(334 \tau ^{16}+42720 \tau ^{14}+654085 \tau ^{12}-770837 \tau ^{10}-17837119 \tau ^8-4216611 \tau ^6+126995832 \tau ^4+515553700 \tau ^2+58320000\right) \omega ^{14}\cr
&+&\tau  \left(-1076 \tau ^{16}+166723 \tau ^{14}+507653 \tau ^{12}-6694665 \tau ^{10}+8248423 \tau ^8+70650686 \tau ^6+250275432 \tau ^4+524896200 \tau ^2+19440000\right) \omega ^{12}\cr
&+&2 \tau ^3 \left(-1350 \tau ^{14}+128734 \tau ^{12}-378886 \tau ^{10}-928564 \tau ^8+28069597 \tau ^6-11328337 \tau ^4+100408896 \tau ^2+88535700\right) \omega ^{10}\bigr)/\cr
&/&\Bigl(72 \tau ^4 \omega ^3 \left(\omega ^2+1\right)^3 \left(\left(\tau ^2-3\right) \omega +4 \tau \right)^2 \left(\left(\tau ^2+5\right) \omega +4 \tau \right) \left(\left(\tau ^2+5\right) \omega ^2+\tau ^2-2 \tau  \omega \right)^3\Bigr),\cr
p_{6}&=&\Bigl(8208 \tau ^{13}+108 \tau ^{12} \left(103 \tau ^2-1293\right) \omega +108 \tau ^{11} \left(52 \tau ^4-1194 \tau ^2+8965\right) \omega ^2+27 \tau ^{10} \left(47 \tau ^6-757 \tau ^4+14396 \tau ^2-157832\right) \omega ^3 \cr
&+&27 \tau ^8 \left(208 \tau ^8-7411 \tau ^6+66461 \tau ^4-181320 \tau ^2-974888\right) \omega ^5\cr
&+&\tau ^6 \left(8019 \tau ^{10}-354942 \tau ^8+845528 \tau ^6+10693801 \tau ^4-51362208 \tau ^2-19364400\right) \omega ^7 \cr
&+&27 \tau ^9 \left(4 \tau ^8+507 \tau ^6-12093 \tau ^4-3668 \tau ^2+470560\right) \omega ^4\cr
&-&4 \tau  \left(\tau ^2+5\right)^2 \left(13 \tau ^{12}-139 \tau ^{10}-729 \tau ^8+6127 \tau ^6+5228 \tau ^4-69060 \tau ^2+93600\right) \omega ^{18}\cr
&+&\tau ^4 \left(270 \tau ^{12}\!-\!221988 \tau ^{10}\!-\!2159450 \tau ^8\!+\!16398269 \tau ^6\!-\!2922643 \tau ^4\!-\!25017876 \tau ^2\!+\!81649620\right) \omega ^9\cr
&+&\tau ^7 \left(594 \tau ^{10}\!-\!16308 \tau ^8\!+\!96579 \tau ^6\!-\!6157663 \tau ^4\!+\!21565656 \tau ^2\!+\!35163288\right) \omega ^6 \cr
&-&4 \left(\tau ^2+5\right) \left(232 \tau ^{14}+959 \tau ^{12}-18025 \tau ^{10}-64807 \tau ^8+321133 \tau ^6+668320 \tau ^4-2272500 \tau ^2-36000\right) \omega ^{17}\cr
&+&2 \tau ^2 \left(-3213 \tau ^{14}-1004 \tau ^{12}+157090 \tau ^{10}-318712 \tau ^8-4085577 \tau ^6+2903420 \tau ^4+37851340 \tau ^2+70475200\right) \omega ^{15}\cr
&+&2 \tau ^2 \left(-6222 \tau ^{14}-5104 \tau ^{12}-172199 \tau ^{10}-1991861 \tau ^8+4135121 \tau ^6+23978589 \tau ^4+116082968 \tau ^2+70105500\right) \omega ^{13}\cr
&+&\tau ^2 \left(-11132 \tau ^{14}-4772 \tau ^{12}-2576993 \tau ^{10}+810934 \tau ^8+38117061 \tau ^6+33242626 \tau ^4+241482060 \tau ^2+46737000\right) \omega ^{11}\cr
&+&\tau ^5 \left(1296 \tau ^{12}-92205 \tau ^{10}+737434 \tau ^8-6748062 \tau ^6-9509285 \tau ^4+69986592 \tau ^2-28888272\right) \omega ^8\cr
&-&2 \tau  \left(120 \tau ^{16}+319 \tau ^{14}+24198 \tau ^{12}+159428 \tau ^{10}-816870 \tau ^8-5762295 \tau ^6+2979880 \tau ^4+39890900 \tau ^2+15684000\right) \omega ^{16} \cr
&-&2 \tau  \left(76 \tau ^{16}+14928 \tau ^{14}+130243 \tau ^{12}-463259 \tau ^{10}-4214011 \tau ^8+1534947 \tau ^6+32861952 \tau ^4+126418300 \tau ^2+14580000\right) \omega ^{14}\cr
&+&\tau  \left(544 \tau ^{16}-81056 \tau ^{14}-406756 \tau ^{12}+3282267 \tau ^{10}-1552982 \tau ^8-32481025 \tau ^6-127535136 \tau ^4-262448100 \tau ^2-9720000\right) \omega ^{12}\cr
&+&\tau ^3 \left(1350 \tau ^{14}-127930 \tau ^{12}+405226 \tau ^{10}+564610 \tau ^8-28835557 \tau ^6+11794651 \tau ^4-100408896 \tau ^2-88535700\right) \omega ^{10}\Bigr)/\cr
&/&\Bigl(72 \tau ^4 \omega ^3 \left(\omega ^2+1\right)^3 \left(\left(\tau ^2-3\right) \omega +4 \tau \right)^2 \left(\left(\tau ^2+5\right) \omega +4 \tau \right) \left(\left(\tau ^2+5\right) \omega ^2+\tau ^2-2 \tau  \omega \right)^3\Bigr),\cr
p_{7}&=&\frac{3}{512} \Biggl(\frac{16 \left(3 \tau ^2-55\right)}{\omega }-\frac{\left(3 \tau ^4+6 \tau ^2-13\right) \left(\tau ^2+5\right)^2}{\left(\tau ^2-3\right)^2 \left(\left(\tau ^2-3\right) \omega +4 \tau \right)}-\frac{6912 \left(\tau ^4+\tau ^2\right)}{\left(\tau ^2-3\right) \left(\left(\tau ^2+5\right) \omega +4 \tau \right)^3}+\frac{16 \left(\tau ^3+\tau \right) \left(\tau ^2+5\right)^2}{\left(\tau ^2-3\right)^2 \left(\left(\tau ^2-3\right) \omega +4 \tau \right)^2}\cr
&-&\frac{144 \tau  \left(7 \tau ^6+5 \tau ^4-11 \tau ^2-73\right)}{\left(\tau ^2-3\right)^2 \left(\left(\tau ^2+5\right) \omega +4 \tau \right)^2} 
-\frac{128 \left(\tau ^6\!-\!21 \tau ^4\!+\!96 \tau ^2\!-\!128\right)}{\tau  \left(\tau ^2-3\right)^2}\cr
&+&\frac{128 \left(\tau ^8-66 \tau ^6+597 \tau ^4-1856 \tau ^2-4 \left(5 \tau ^6\!-\!62 \tau ^4\!+\!217 \tau ^2\!-\!240\right) \tau  \omega +1920\right)}{\left(\tau ^2-3\right)^3 \left(4 \tau  \omega ^2+\tau +8 \omega \right)} \cr
&-&\frac{45 \tau ^{10}+157 \tau ^8+306 \tau ^6-3414 \tau ^4-943 \tau ^2+13929}{\left(\tau ^2-3\right)^3 \left(\left(\tau ^2+5\right) \omega +4 \tau \right)}+\frac{320 \tau }{\omega ^2}\Biggr),\cr
\end{eqnarray*}
\begin{eqnarray*}
q_{1}&=&\frac{-4 \tau ^3+8 \tau  \left(\tau ^2+5\right) \omega ^4+3 \tau  \left(14-5 \tau ^2\right) \omega ^2-\tau ^2 \left(\tau ^2+18\right) \omega +\left(-3 \tau ^4+29 \tau ^2+70\right) \omega ^3}{\tau  \omega ^3 \left(\left(\tau ^2+5\right) \omega +4 \tau \right) \left(4 \tau  \omega ^2+\tau +8 \omega \right)},\cr
q_{2}&=&\frac{4 \tau ^3+\left(\tau ^2+3\right) \left(3 \tau ^2-10\right) \omega ^3+\tau  \left(15 \tau ^2+14\right) \omega ^2+\tau ^2 \left(\tau ^2+10\right) \omega -8 \tau  \omega ^4}{4 \omega  \left(\omega ^2+1\right) \left(\left(\tau ^2-3\right) \omega +4 \tau \right) \left(\left(\tau ^2+5\right) \omega ^2+\tau ^2-2 \tau  \omega \right)},\cr
q_{3}&=&\Bigl(9 \Bigl(16 \tau ^6+4 \left(5 \tau ^2-39\right) \tau ^5 \omega +48 \left(\tau ^4+3 \tau ^2-6\right) \omega ^6+4 \left(2 \tau ^4-30 \tau ^2+85\right) \tau ^4 \omega ^2+\left(\tau ^8-4 \tau ^6-89 \tau ^4+780\right) \tau  \omega ^5 \cr
&+&\left(5 \tau ^6+41 \tau ^4-292 \tau ^2-428\right) \tau ^2 \omega ^4+\left(\tau ^6-19 \tau ^4+236 \tau ^2-264\right) \tau ^3 \omega ^3\Bigr)\Bigr)/\cr
&/&\Bigl(16 \tau  \omega ^3 \left(\left(\tau ^2-3\right) \omega +4 \tau \right)^2 \left(\left(\tau ^2+5\right) \omega +4 \tau \right) \left(\left(\tau ^2+5\right) \omega ^2+\tau ^2-2 \tau  \omega\right) \Bigr),\cr
 q_{4}&=&\frac{4 \tau ^4-8 \tau  \left(\tau ^2+6\right) \omega ^5+\tau ^2 \left(\tau ^2+4\right) \left(\tau ^2+16\right) \omega ^2+\tau  \left(15 \tau ^4+32 \tau ^2-4\right) \omega ^3+4 \tau ^3 \left(\tau ^2-1\right) \omega +3 \left(\tau ^6+3 \tau ^4+4 \tau ^2-20\right) \omega ^4}{4 \tau ^2 \omega ^2 \left(\omega ^2+1\right) \left(\left(\tau ^2-3\right) \omega +4 \tau \right) \left(\left(\tau ^2+5\right) \omega ^2+\tau ^2-2 \tau  \omega \right)},\cr
q_{5}&=&\Bigl(864 \tau ^{13}+216 \tau ^{12} \left(5 \tau ^2-183\right) \omega +216 \tau ^{11} \left(2 \left(\tau ^2-97\right) \tau ^2+1733\right) \omega ^2 \cr
&+&54 \tau ^{10} \left(\tau ^6-259 \tau ^4+3488 \tau ^2-36196\right) \omega ^3-54 \tau ^9 \left(23 \tau ^6+1627 \tau ^4+2340 \tau ^2-125600\right) \omega ^4\cr
&-&\tau  \left(25692274 \tau ^6+83535672 \tau ^4+204492600 \tau ^2+3 \left(7523 \tau ^6-25795 \tau ^4-960929 \tau ^2+458955\right) \tau ^8+9720000\right) \omega ^{12}\cr
&+&54 \tau ^8 \left(\tau ^8-1331 \tau ^6+14776 \tau ^4-39404 \tau ^2-296152\right) \omega ^5 +4 \tau  \left(\tau ^2+5\right)^2 \left(159 \tau ^{10}+338 \tau ^8-4945 \tau ^6+2276 \tau ^4+57840 \tau ^2-95400\right) \omega ^{18} \cr
&-&2 \tau ^6 \left(405 \tau ^{10}+56997 \tau ^8-268604 \tau ^6-3107148 \tau ^4+15966504 \tau ^2+10198008\right) \omega ^7\cr
&+&2 \tau ^7 \left(-6966 \tau ^8+56295 \tau ^6-1415779 \tau ^4+6023376 \tau ^2+12630168\right) \omega ^6\cr
&-&2 \tau ^4 \left(1485 \tau ^{12}+26491 \tau ^{10}+331952 \tau ^8-5295916 \tau ^6+618370 \tau ^4+16582212 \tau ^2-26011260\right) \omega ^9\cr
&-&2 \tau ^5 \left(17631 \tau ^{10}-219662 \tau ^8+1534098 \tau ^6+2355751 \tau ^4-26064072 \tau ^2+4701456\right) \omega ^8 \cr
&-&\left(\tau ^2+5\right) \left(259 \tau ^{14}-2712 \tau ^{12}-59326 \tau ^{10}-66192 \tau ^8+1232455 \tau ^6+1284560 \tau ^4-8273700 \tau ^2-72000\right) \omega ^{17}\cr
&+&\tau  \left(2303 \tau ^{14}+5324 \tau ^{12}+70078 \tau ^{10}+1821060 \tau ^8+6567967 \tau ^6-11641720 \tau ^4-63630500 \tau ^2-30084000\right) \omega ^{16}\cr
&+&\tau ^2 \left(-1591 \tau ^{14}+22035 \tau ^{12}+265876 \tau ^{10}-297240 \tau ^8-4586381 \tau ^6+6349577 \tau ^4+60686600 \tau ^2+125583700\right) \omega ^{15}\cr
&-&\tau  \left(3602 \tau ^{14}+1623 \tau ^{12}-1226535 \tau ^{10}-5967603 \tau ^8+5284837 \tau ^6+59228208 \tau ^4+199074500 \tau ^2+29160000\right) \omega ^{14}\cr
&+&\tau ^2 \left(-3729 \tau ^{14}+49407 \tau ^{12}+64261 \tau ^{10}-1452345 \tau ^8+8741292 \tau ^6+39545982 \tau ^4+162444348 \tau ^2+125631000\right) \omega ^{13}\cr
&+&\tau ^2 \left(-4589 \tau ^{14}+39785 \tau ^{12}-847581 \tau ^{10}+3094217 \tau ^8+27647442 \tau ^6+15692974 \tau ^4+158351760 \tau ^2+41877000\right) \omega ^{11}\cr
&-&2 \tau ^3 \left(20512 \tau ^{12}-196082 \tau ^{10}-461212 \tau ^8+8120773 \tau ^6-5856249 \tau ^4+23546808 \tau ^2+34406100\right) \omega ^{10}\Bigr)/\cr
&/&\Bigl(72 \tau ^4 \omega ^5 \left(\omega ^2+1\right)^3 \left(\left(\tau ^2-3\right) \omega +4 \tau \right)^2 \left(\left(\tau ^2+5\right) \omega +4 \tau \right) \left(\left(\tau ^2+5\right) \omega ^2+\tau ^2-2 \tau  \omega \right)^3\Bigr),\cr
q_{6}&=&\Bigl(-432 \tau ^{13}+108 \tau ^{12} \left(183-5 \tau ^2\right) \omega -108 \tau ^{11} \left(2 \left(\tau ^2-97\right) \tau ^2+1733\right) \omega ^2-27 \tau ^{10} \left(\tau ^6-259 \tau ^4+3488 \tau ^2-36196\right) \omega ^3 \cr
&+&27 \tau ^9 \left(23 \tau ^6+1627 \tau ^4+2340 \tau ^2-125600\right) \omega ^4 \cr
&+&\tau  \left(13006691 \tau ^6+40454592 \tau ^4+102246300 \tau ^2+3 \left(3736 \tau ^6-20672 \tau ^4-436699 \tau ^2+506754\right) \tau ^8+4860000\right) \omega ^{12} \cr
&-&27 \tau ^8 \left(\tau ^8-1331 \tau ^6+14776 \tau ^4-39404 \tau ^2-296152\right) \omega ^5\cr
&-&16 \tau  \left(\tau ^2+5\right)^2 \left(21 \tau ^{10}+37 \tau ^8-647 \tau ^6+403 \tau ^4+7410 \tau ^2-12600\right) \omega ^{18} \cr
&+&\tau ^6 \left(405 \tau ^{10}+56997 \tau ^8-267932 \tau ^6-3106860 \tau ^4+15966504 \tau ^2+10198008\right) \omega ^7 \cr
&+&\tau ^7 \left(6966 \tau ^8-56295 \tau ^6+1416163 \tau ^4-6023376 \tau ^2-12630168\right) \omega ^6\cr
&+&\tau ^4 \left(1485 \tau ^{12}+26641 \tau ^{10}+329642 \tau ^8-5370514 \tau ^6+566968 \tau ^4+16582212 \tau ^2-26011260\right) \omega ^9\cr
&+&\tau ^5 \left(17631 \tau ^{10}-219206 \tau ^8+1533090 \tau ^6+2329423 \tau ^4-26064072 \tau ^2+4701456\right) \omega ^8\cr
&+&2 \left(\tau ^2+5\right) \left(73 \tau ^{14}-912 \tau ^{12}-15454 \tau ^{10}-9672 \tau ^8+321793 \tau ^6+271400 \tau ^4-2144700 \tau ^2+36000\right) \omega ^{17}\cr
&-&2 \tau  \left(347 \tau ^{14}+2522 \tau ^{12}+38752 \tau ^{10}+470358 \tau ^8+1360273 \tau ^6-3385600 \tau ^4-15419900 \tau ^2-6828000\right) \omega ^{16} \cr 
&+&2 \tau ^2 \left(379 \tau ^{14}-3036 \tau ^{12}-65188 \tau ^{10}-37326 \tau ^8+896873 \tau ^6-1031582 \tau ^4-13813040 \tau ^2-31431400\right) \omega ^{15} \cr
&+&8 \tau  \left(137 \tau ^{14}+3513 \tau ^{12}-65535 \tau ^{10}-441378 \tau ^8+105220 \tau ^6+3742965 \tau ^4+12729050 \tau ^2+1822500\right) \omega ^{14}\cr
&+&\tau ^2 \left(1860 \tau ^{14}-30216 \tau ^{12}+50329 \tau ^{10}+1130583 \tau ^8-4873917 \tau ^6-22343607 \tau ^4-81935892 \tau ^2-62815500\right) \omega ^{13}\cr
&+&\tau ^2 \left(2296 \tau ^{14}-20398 \tau ^{12}+365475 \tau ^{10}-1463491 \tau ^8-13007025 \tau ^6-7423697 \tau ^4-79175880 \tau ^2-20938500\right) \omega ^{11}\cr
&+&\tau ^3 \left(20536 \tau ^{12}-197678 \tau ^{10}-549022 \tau ^8+8092717 \tau ^6-5553219 \tau ^4+23546808 \tau ^2+34406100\right) \omega ^{10}\Bigr)/\cr
&/&\Bigl(72 \tau ^4 \omega ^5 \left(\omega ^2+1\right)^3 \left(\left(\tau ^2-3\right) \omega +4 \tau \right)^2 \left(\left(\tau ^2+5\right) \omega +4 \tau \right) \left(\left(\tau ^2+5\right) \omega ^2+\tau ^2-2 \tau  \omega \right)^3\Bigr),\cr
q_{7}&=&\Bigl(3 \Bigl(-256 \tau ^8-4 (\tau -2) \tau  (\tau +2) \left(\tau ^2+5\right)^3 \left(3 \tau ^2-8\right) \omega ^9+128 \tau ^7 \left(1-2 \tau ^2\right) \omega -16 \tau ^6 \left(6 \tau ^4+22 \tau ^2-895\right) \omega ^2\cr
&+&\left(\tau ^2+5\right)^2 \left(7 \tau ^8-286 \tau ^6+667 \tau ^4+1072 \tau ^2-1280\right) \omega ^8 +\tau  \left(\tau ^2+5\right) \left(153 \tau ^8-2246 \tau ^6-4159 \tau ^4+15440 \tau ^2+4480\right) \omega ^7 \cr
&-&\tau ^4 \left(\tau ^8+178 \tau ^6-17295 \tau ^4+1872 \tau ^2+70368\right) \omega ^4-8 \tau ^5 \left(2 \tau ^6+55 \tau ^4-3043 \tau ^2-246\right) \omega ^3 \cr
&-&2 \tau ^2 \left(\tau ^{10}-688 \tau ^8+2365 \tau ^6+28110 \tau ^4+13224 \tau ^2-41320\right) \omega ^6-\tau ^3 \left(31 \tau ^8-6530 \tau ^6+6567 \tau ^4+99816 \tau ^2+27472\right) \omega ^5\Bigr)\Bigr)/\cr
&/&\Bigl(2 \tau  \omega ^4 \left(\left(\tau ^2-3\right) \omega +4 \tau \right)^2 \left(\left(\tau ^2+5\right) \omega +4 \tau \right)^3 \left(4 \tau  \omega ^2+\tau +8 \omega \right)\Bigr)\,.
\end{eqnarray*}

\begin{figure}[tbp]
	\centering	
	\includegraphics[width=0.65\columnwidth]{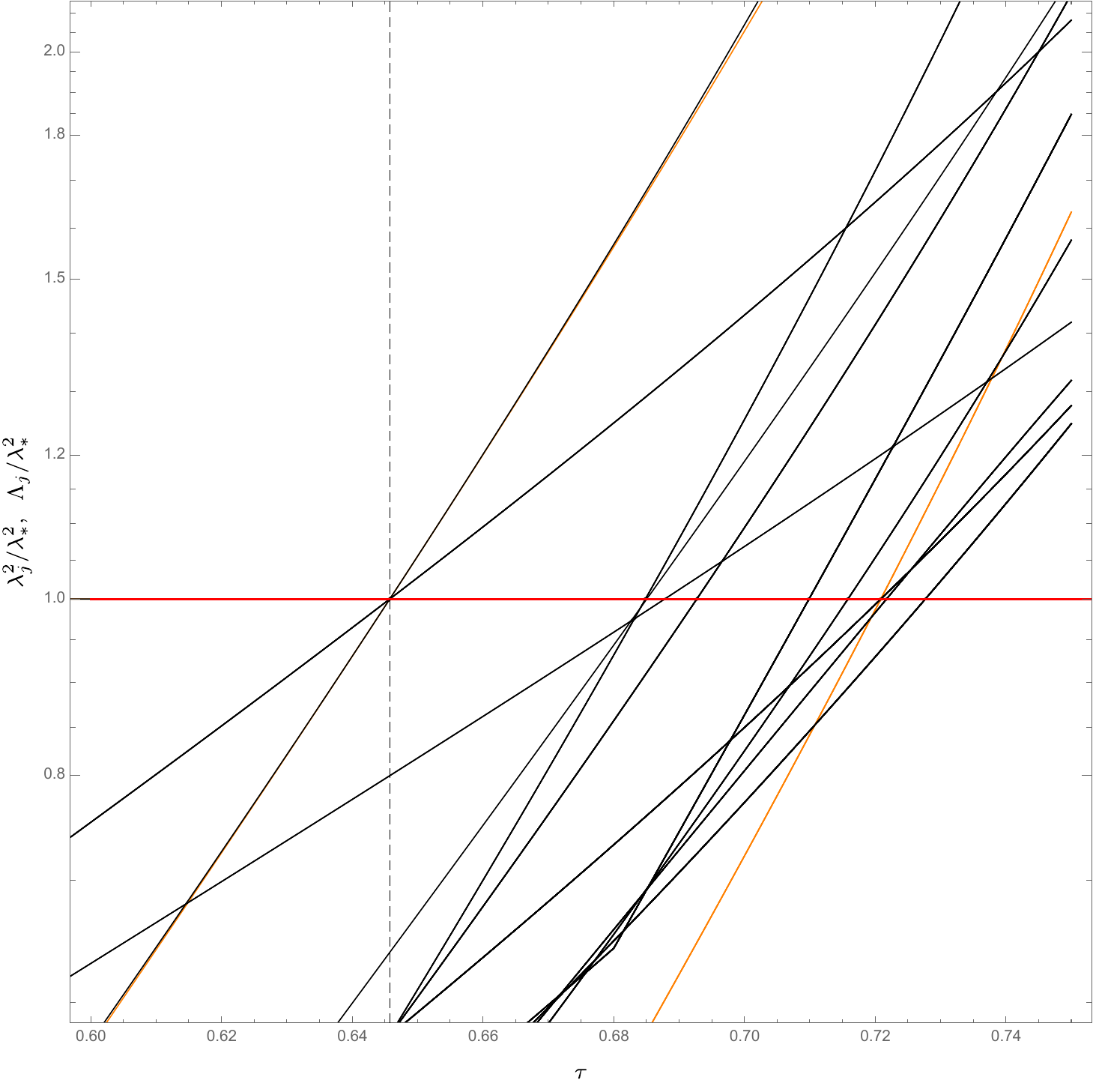}
	\caption{The flow of square eigenvalues $\lambda^2_j(\tau)$ (orange) of $A_0$ (\ref{eq:AA}) and largest 64 eigenvalues -- with degeneracies -- of $\mathbb T$ (\ref{HST}) (black) relative to $\lambda_*^2(\tau)$ (red line), while fixing
	$\omega=1$.
	Note a tiny gap between $\lambda_0^2/\lambda_*^2$ and $\Lambda_0/\lambda_*^2$ for $\tau>\tau_{\rm c}=0.6458$.}
	\label{fig:LF}
\end{figure}
 \noindent
Explicit expression for {\em Drude weight} (\ref{Drude}) of boundary operator $B=\frac{1}{\sqrt{3}}\vec{\sigma_1}\cdot\vec{\sigma_2}$, under condition $\Lambda_0=\lambda_0^2$, evaluates to:
\begin{eqnarray}
D&=&\Bigl(\omega ^2 \bigl(16 \tau ^5+20 \left(\tau ^2-5\right) \tau ^4 \omega +2 \left(-21 \tau ^6-18 \tau ^4+145 \tau ^2+150\right)
   \tau  \omega ^4+\left(\tau ^6-154 \tau ^4-235 \tau ^2+160\right) \tau ^2 \omega ^3 \nonumber \cr
   &+&4 \left(2 \tau ^4-55 \tau ^2-65\right) \tau
   ^3 \omega ^2-\left(4 \tau ^8-\tau ^6-50 \tau ^4+15 \tau ^2+180\right) \omega ^5\bigr)\Bigr)/\Bigl(3 (\tau\!-\!\omega )^2 \left(\left(\tau^2\!-\!4\right) \omega +4 \tau \right) \left(\left(4 \tau ^2\!+\!9\right) \omega ^2+\tau ^2+6 \tau  \omega \right)^2\Bigr) \nonumber \\
   &-& r \frac{4 \omega  \left(\left(\tau ^2+3\right) \omega +2 \tau \right) \left(\left(\tau ^4-6\right) \omega ^2+5 \left(\tau ^2+1\right)
   \tau  \omega +5 \tau ^2\right)}{3 (\tau -\omega ) \left(\left(\tau ^2-4\right) \omega +4 \tau \right) \left(\left(4 \tau
   ^2+9\right) \omega ^2+\tau ^2+6 \tau  \omega \right)^2}\,. \label{eq:Drude1}
\end{eqnarray}

\noindent
It is useful to separately express the Hilbert-Schmidt square-norm of QLEM (which simplifies a bit using $SO(3)$ symmetry):
\be
(Q|Q) = \frac{\sum_{\nu=0}^3 (\bra{a^\nu}\otimes\bra{a^\nu})\ket{\rm R}}{(\bra{r_*}\otimes \bra{r_*})\ket{\rm R}} = \frac{\bra{a^1}A^1\ket{r_*}^2}{\lambda_*^2 D}.
\label{eq:QQ}
\ee}
\end{widetext}
 
\end{document}